\documentclass[pra,showpacs,eqsecnum,twocolumn]{revtex4}

\usepackage{graphicx}
\usepackage{amsmath}
\usepackage[urlcolor=blue, hyperindex, colorlinks, bookmarks=true,linkcolor=black,citecolor=black]{hyperref}
\begin{document}

\newcommand{\sch}{Schr\"odinger }
\newcommand{\schs}{Schr\"odinger's }
\newcommand{\beq}{\begin{equation}}
\newcommand{\eeq}{\end{equation}}
\newcommand{\bqa}{\begin{eqnarray}}
\newcommand{\eqa}{\end{eqnarray}}
\newcommand{\nn}{\nonumber}
\newcommand{\nl}{\nn \\ &&}
\newcommand{\dg}{^\dagger}
\newcommand{\rt}[1]{\sqrt{#1}\,}
\newcommand{\smallfrac}[2]{\mbox{$\frac{#1}{#2}$}}
\newcommand{\bra}[1]{\langle{#1}|}
\newcommand{\ket}[1]{|{#1}\rangle}
\newcommand{\bl}{{\Bigl(}}
\newcommand{\br}{{\Bigr)}}
\newcommand{\erf}[1]{Eq.~(\ref{#1})}
\newcommand{\erfs}[2]{Eqs.~(\ref{#1}) and (\ref{#2})}
\newcommand{\erft}[2]{Eqs.~(\ref{#1}) -- (\ref{#2})}
\newcommand{\lfrac}[2]{({#1})/{#2}}

\newcommand{\ak}{\ket{\{a_{k}\}}}
\newcommand{\ab}{\bra{\{a_{k}\}}}
\newcommand{\qk}{\ket{\{q_{k}\}}}
\newcommand{\qb}{\bra{\{q_{k}\}}}
\newcommand{\psiqk}[1]{\ket{{\psi}_{\{q_{k}\}}({#1})}}
\newcommand{\psiqb}[1]{\bra{{\psi}_{\{q_{k}\}}({#1})}}
\newcommand{\lpsiqk}[1]{\ket{\tilde{\psi}_{\{q_{k}\}}({#1})}}
\newcommand{\lpsiqb}[1]{\bra{\tilde{\psi}_{\{q_{k}\}}({#1})}}

\newcommand{\psizk}[1]{\ket{\psi_{z}({#1})}}
\newcommand{\psizb}[1]{\bra{\psi_{z}({#1})}}
\newcommand{\lpsizk}[1]{\ket{\tilde{\psi}_{z}({#1})}}
\newcommand{\lpsizb}[1]{\bra{\tilde{\psi}_{z}({#1})}}

\newcommand{\f}[1]{\hspace{-1mm}~^{(#1)}\hspace{-.7mm}\hat{f}}
\newcommand{\F}[1]{\hspace{-1mm}~^{(#1)}\hspace{-.7mm}\hat{F}}
\newcommand{\q}[1]{\hspace{-1mm}~^{(#1)}\hspace{-.5mm}\hat{q}}
\newcommand{\Q}[1]{\hspace{-1mm}~^{(#1)}\hspace{-.5mm}\hat{Q}}
\newcommand{\fu}[1]{\hspace{-1mm}~^{(#1)}\hspace{-.7mm}{f}}
\newcommand{\Fu}[1]{\hspace{-1mm}~^{(#1)}\hspace{-.7mm}{F}}
\newcommand{\qu}[1]{\hspace{-1mm}~^{(#1)}\hspace{-.5mm}{q}}
\newcommand{\Qu}[1]{\hspace{-1mm}~^{(#1)}\hspace{-.5mm}{Q}}

\title{A perturbative approach to non-Markovian stochastic \sch equations}
\date{\today}
\author{Jay Gambetta}
\affiliation{Centre for Quantum Dynamics, School of Science,
Griffith University, Brisbane 4111, Australia}
\author{H. M. Wiseman} \email{h.wiseman@gu.edu.au}
\affiliation{Centre for Quantum Dynamics, School of Science,
Griffith University, Brisbane 4111, Australia}

\begin{abstract}
In this paper we present a perturbative procedure that allows one to numerically solve diffusive non-Markovian Stochastic \sch equations, for a wide range of memory functions. To illustrate this procedure numerical results are presented for a classically driven two level atom immersed in a environment with a simple memory function. It is observed that as the order of the perturbation is increased the numerical results for the ensembled average state $\rho_{\rm red}(t)$ approach the exact reduced state found via Imamo\=glu's enlarged system method [Phys. Rev. A. {\bf 50}, 3650 (1994)].
\end{abstract}

 \maketitle

\section{Introduction}

A common problem in physics is to model open quantum systems. They consists of a small
system immersed in a bath (environment). Due to the large Hilbert space of the bath it is
convenient to describe the system by its reduced state. The reduced state is defined as
\begin{equation}\label{ReducedState} \rho_{\rm red}(t)={\rm Tr}_{\rm
bath}[\ket{\Psi(t)}\bra{\Psi(t)}]. \end{equation} where $\ket{\Psi(t)}$ is the combined
system state, found from the \sch equation for the open quantum system.

It has been shown \cite{Nak58,Zwa60} by a projection-operator
method that we can write a general master equation for the reduced
state as
\begin{equation} \label{Master} \dot{\rho}_{\rm
red}(t)=-\frac{i}{\hbar}[\hat{H}(t),\rho_{\rm
red}(t)]+\int_{0}^{t}{\cal K}(t,s)[\hat{L}]\rho_{\rm red}(s) ds
\end{equation} where ${\cal K}(t,s)[\hat{L}]$ is the `memory time'
superoperator. It (operators on) the system operator $\hat{L}$ and
represents how the bath affects the system. The problem with this
equation is that in general ${\cal K}(t,s)[\hat{L}]$ can not be
explicitly evaluated.

The most notable approximation used is the Born-Markov one. This
arises when the environmental influences on the system are
instantaneous. Mathematical consistency requires that this results
in a Lindblad master equation, of the form \cite{Lin76}
\begin{equation}\label{MarkovReducedState} \dot\rho_{\rm red} (t)
=-\frac{i}{\hbar}[\hat{H}(t),\rho_{\rm red}(t)]+\gamma {\cal
D}[\hat{L}]\rho_{\rm red} (t) ,
\end{equation} where ${\cal D}[\hat{L}]$ is the superoperator that represent the damping
of the system into the bath. It has the form
\begin{equation}\label{DOperator} {\cal D}[\hat{L}]\rho_{\rm
red}=\hat{L}\rho_{\rm red}\hat{L}\dg-\smallfrac12 \hat{L}\dg
\hat{L}\rho_{\rm red}-\smallfrac12 \rho_{\rm
red}\hat{L}\dg\hat{L}. \end{equation} This equation can be solved
deterministically \cite{GarZol00} or by the stochastic \sch
approach \cite{GarZol00,DalCasMol92,GarParZol92,Car93}.

For the non-Markovian situation there have been many attempts at
finding solutions to \erf{Master}. However, some have the problem
that it is hard to ensure the positivity requirement on $\rho_{\rm
red}(t)$ \cite{Red65}. A method that does ensure the positivity
requirement on the reduced state is the non-Markovian stochastic
\sch equation (SSE) approach
\cite{Dio96,DioStr97,DioGisStr98,StrDioGis99,Str01,Cre00,Bud00,GamWis02}.
A non-Markovian SSE generates stochastic pure states
${\ket{\psi_{z}(t)}}$ that should
satisfy\begin{equation}\label{EnsembleNonMarkov} \rho_{\rm
red}(t)=E[{\ket{\psi_{z}(t)}}{\bra{\psi_{z}(t)}}],
\end{equation} where $z(t)$ is some noise function which is non-white
and $E$ denotes the ensemble average over $z(t)$. To solve these
non-Markovian SSE one has to take into account the past behavior
of the system and bath, giving rise to a functional derivative in
the attempt to derive a SSE. This presents a problem as for most
systems an exact solution to the functional derivative does not
exists. Thus at present an exact non-Markovian SSE only exists for
simple systems, which can be solve exactly via other methods, like
the undriven two level atom (TLA). For this and more examples see
Ref. \cite{DioGisStr98,GamWis02}.

Recently Yu, Di\'osi, Gisin and Strunz (YDGS) have developed
explicitly a `post-Markovian' perturbation method to first order
that allows solutions for systems that are close to the Markovian
limit \cite{YuDioGisStr99,YuDioGisStr00}. In this paper we present
a perturbation method that can be carried to arbitrary order and
so is not limited to the post Markovian regime. However we must
place a requirement on the form of the memory functions. This
requirement is that the memory function must take the form
\begin{equation}\label{memory} \alpha(t-s)=\sum_{j=1}^{J}|G_{j}|^2
e^{-\kappa_{j}|t-s|/2-i(\omega_{j}-\Omega)(t-s)},
\end{equation}for some finite (and, in practice, relatively small) $J$. It should be
noted also that we have not proven convergence of our perturbation
theory and this theory is only valid for a zero-temperature bath.

The format of this paper is as follows. In Sec. \ref{NMSSE} we
present a general outline of the theory of non-Markovian SSE. This
is basically a summary of the results of Refs.
\cite{Dio96,DioStr97,DioGisStr98,StrDioGis99,GamWis02}. In Sec.
\ref{Pert} our perturbation method is presented. In Secs.
\ref{Enl} we outline Imamo${\rm \bar{g}}$lu enlarged system method
\cite{Ima94,SteIma96}. In Sec. \ref{TLAapply} we apply our
perturbation method to a simple system, a driven TLA and compare
our results for $\rho_{\rm red}(t)$ with the enlarged system
methods. In Sec. \ref{YDGS} we investigate YDGS post-Markovian
perturbation method \cite{YuDioGisStr99,YuDioGisStr00}. Finally we
conclude with a discussion of the potential applications of our
results in Sec. \ref{Conclude}.

\section{Non-Markovian Stochastic \sch Equations} \label{NMSSE}

In this section we will present an outline of the theory we
presented in \cite{GamWis02}, which is an extension of Di\'osi,
Gisin and Strunz (DGS) diffusive Non-Markovian SSEs
\cite{Dio96,DioStr97,DioGisStr98,StrDioGis99} which allows for
real-valued noise $z(t)$.

\subsection{Underlying Dynamics} \label{dynamics}

The non-Markovian SSEs developed in references
\cite{GamWis02,Dio96,DioStr97,DioGisStr98,StrDioGis99} are valid when the dynamics of the
open quantum system can be described by the total Hamiltonian \begin{equation}
\label{HamiltonianTotal} \hat{H}_{\rm tot}=\hat{H}_{\rm
sys}\otimes\hat{1}+\hat{1}\otimes\hat{H}_{\rm bath}+\hat{V}.
\end{equation} The system Hamiltonian is $\hat{H}_{\rm sys}=\hat{H}_{\Omega}+\hat{H}$.
The bath is modelled by a collection of harmonic oscillators, so the Hamiltonian for the
bath is
\begin{equation}\label{HamiltonianBath} \hat{H}_{\rm
bath}=\hbar\sum_{k}\omega_{k}\hat{a}_{k}\dg \hat{a}_{k}, \end{equation} where
$\hat{a}_{k}$ and $\omega_{k}$ are the lowering operator and angular frequency of the
$k^{\rm th}$ mode respectively. This is the standard model for the electromagnetic field.
The interaction Hamiltonian has the form
\begin{equation}\label{HamiltonianInteraction}
\hat{V}=i\hbar(\hat{L}\hat{b}\dg-\hat{L}\dg\hat{b}), \end{equation} where we have defined
the bath lowering operators $\hat{b}$ as $\hat{b}=\sum_{k}g_{k}\hat{a}_{k}$. That is, the
coupling amplitude of the $k^{\rm th}$ mode to the system is $g_{k}$.

For calculation purposes we define the non-Markovian SSE in an interaction picture. This allows us to move the fast
dynamics placed on the state by the Hamiltonians $\hat{H}_{\Omega}$ and $\hat{H}_{\rm bath}$ to the operators. The
unitary evolution operator for this transformations is \begin{equation}\label{UnitaryFree}
U(t,0)=e^{-\frac{i}{\hbar}(\hat{H}_{\Omega}\otimes\hat{1}+\hat{1}\otimes\hat{H}_{\rm bath})(t-0)}.
\end{equation} Thus the combined state in the interaction picture is define as
\begin{equation}\label{UnitarySch} \ket{\Psi(t)}=U\dg(t,0)\ket{\Psi(t)_{\rm Sch}},
\end{equation} and an arbitrary operator $\hat{A}$ becomes \begin{equation} \hat{A}_{\rm
int}=U\dg(t,0)\hat{A}U(t,0). \end{equation} This allows us to write the \sch equation as
\begin{equation} \label{IntSchEquation} d_t\ket{{\Psi}(t)} =-\frac{i}{\hbar}[ \hat{H}_{\rm
int}(t) + \hat{V}_{\rm int}(t) ] \ket{\Psi(t)}, \end{equation}  where the Hamiltonians are
\begin{eqnarray} \label{HamiltonianExtraInt} \hat{H}_{\rm int}(t)&=&
U\dg(t,0)\hat{H}U(t,0), \end{eqnarray} and \begin{eqnarray} \label{HamiltonianInteractionInt} \hat{V}_{\rm int}(t)&=&
i\hbar[  \hat{L}e^{-i\Omega t}\hat{b}_{\rm int}\dg(t) -\hat{L}\dg e^{i\Omega t}\hat{b}_{\rm int}(t) ], \end{eqnarray}
where \begin{eqnarray} \label{BathOperatorInt} \hat{b}_{\rm int}(t)&=&\sum_{k}g_{k}\hat{a}_{k}e^{-i\omega_{k} t}.
\end{eqnarray} Here we have finally restricted the form of $\hat{H}_{\Omega}$ to be such that $\hat{L}$ in the
interaction picture simply rotates in the complex plane at frequency $\Omega$. That is
$\hat{L}_{\rm int}(t)= \hat{L}e^{-i\Omega t}$.

\subsection{Non-Markovian SSE-Defined}

A non-Markovian SSE is a stochastic differential equation for the system state vector
$\psizk{t}$ containing some non-white noise $z(t)$. It has the property that when
$\psizk{t}\psizb{t}$ is averaged over all possible $z(t)$ one obtains $\rho_{\rm red}$(t).
It should be noted that for a single $\rho_{\rm red}(t)$, $z(t)$ can take many different
functional forms, and we label these different forms as stochastic unravelings
\cite{GamWis02}.

In Ref. \cite{GamWis02} we showed that non-Markovian SSEs can be
derived from quantum measurement theory (QMT), where the different
unravelings correspond to different measurements on the bath. The
two unravelings we considered were the `coherent' or `DGS'
\cite{Dio96,DioStr97,DioGisStr98,StrDioGis99} unraveling and the
`quadrature' unraveling. A special case of our quadrature
unraveling was published in Ref. \cite{BasGhi02}.

As in the Markov limit we can define (at least) two non-Markovian
SSEs, for each unraveling: one for $z(t)$ chosen from an
ostensible distribution (a guessed distribution) and the other for
its actual distribution. The former gives a non-Markovian SSE
linear in the unnormalised state $\lpsizk{t}$, while the latter
gives a non-Markovian SSE non-linear in the normalized state
$\psizk{t}$. In Ref. \cite{GamWis02} we came to the conclusion
that the solution of the actual non-Markovian SSE at time $t$
gives the state the system will be in if a measurement of the bath
is performed at that time. Unlike in the Markov case, linking of
the states through time to make a trajectory turns out to be a
convenient fiction. However, it has been suggested that such
trajectories can be given an interpretation within a non-standard
QMT \cite{Lou01,BarLou02}.

\subsubsection{Coherent Unravelling-Outlined}

The first unravelling we consider is the `coherent' unravelling. This unravelling arises
when the bath is projected into a coherent state. We define the coherent state as
\begin{equation} \ak=\prod_{k} \frac{1}{\rt{\pi}}e^{-|a_{k}|^{2}/2}\sum_{n_{k}}
\frac{a_{k}^{n_{k}}}{\rt{n_{k}!}}\ket{n_{k}},
\end{equation} so that $\hat{1}=\prod_{k}\int\ket{a_{k}}\bra{a_{k}} d^{2}a_{k}$.

In a measurement we can define an operator for the measurement process, the noise
operator. For this measurement it must have the coherent basis as its eigenstate, so the
noise operator is \begin{equation} \label{CNoiseOperator} \hat{z}(t)=\hat{b}_{\rm
int}(t)e^{i\Omega t}=\sum_{k}g_{k}\hat{a}_{k}e^{-i\Omega_{k} t},
\end{equation} where $\Omega_{k}=\omega_{k}-\Omega$. The noise function (eigenvalue of the noise operator) is
\begin{equation} \label{CNoiseFunction} {z}(t)=\sum_{k}g_{k}{a}_{k}e^{-i\Omega_{k} t}.
\end{equation} where ${a}_{k}$ are the results of the projection in the coherent basis.

If we assume an ostensible distribution for ${a}_{k}$ as being the overlap of the coherent state with vacuum state, that
is, it has the form \begin{equation} \label{CProbability} {\Lambda}(\{a_{k}\})=\langle{\{0_k\}}\ak\ab{\{0_k\}}\rangle
={\pi^{-K}}{e^{-\sum_{k}|a_{k}|^{2}}}, \end{equation} where $K=\sum_{k}$. With this ostensible distribution the noise
function has the following correlations
\begin{subequations} \begin{eqnarray} \label{CCorrelation1} \tilde{ E}[{z}(t)z^{*}(s)]&=&\alpha{(t-s)},\\ \label{CCorrelation2}
\tilde{ E}[{z}(t)z(s)]&=&0. \end{eqnarray}
\end{subequations} where the tilde above the $E$ refers to a
average over the ostensible distribution. In \erf{CCorrelation1}
we have defined $\alpha(t-s)$, this function we label the memory
function. On a microscopic level it has the form
\begin{equation}\label{MemoryFunction}
\alpha{(t-s)}=\sum_{k}|g_{k}|^{2}e^{-i\Omega_{k} (t-s)}.
\end{equation}

Using the above ostensible distribution we can define a linear conditional system state as
\begin{equation}
  \lpsizk{t}=\frac{\ab\Psi(t)\rangle}{\rt{{\Lambda}(\{a_{k}\})}}.
\end{equation} Taking the time derivative and using \erf{IntSchEquation}
we get a linear differential equation for $\lpsizk{t}$ of the from \begin{eqnarray}
\label{CLSSEFunctionalForm}
\partial_{t}\lpsizk{t}&=&\Big{[} \frac{-i}{\hbar}\hat{H}_{\rm
int}(t)+z^{*}(t)\hat{L}-\hat{L}\dg
\int_{0}^{t}\alpha{(t-s)}\nl\times\frac{\delta}{\delta z^{*}(s)}
ds\Big{]}\lpsizk{t},
\end{eqnarray} where ${\delta}/{\delta z^{*}(s)}$ represents a functional derivative. For a derivation of this
equation see Ref. \cite{DioStr97,GamWis02}. The functional
derivative in this equation stops us from calling this equation a
non-Markovian SSE, as it means that $\partial_{t}\lpsizk{t}$ does
not depend upon the state $\lpsizk{t}$ at all times for a single
function $z(t)$, but rather also upon states for other noise
functions. That is, we cannot stochastically choose $z(t)$ in
order to generate a trajectory independent of other trajectories.
Instead, all possible trajectories would have to be calculated in
parallel, which in calculation terms amounts to solving the
complete \sch equation \erf{IntSchEquation}. However, as explained
in Refs. \cite{GamWis02,DioGisStr98,StrDioGis99} if we can make
the following ansatz
\begin{equation}\label{AnsatzO} \frac{\delta}{\delta
z^{*}(s)}\lpsizk{t}=\f{0}_{z}(t,s)\lpsizk{t}, \end{equation} then we can write a linear non-Markovian SSE as
\begin{equation} \label{CLSSEOperatorForm}
\partial_{t}\lpsizk{t}=\Big{[} \frac{-i}{\hbar}\hat{H}_{\rm
int}(t)+z^{*}(t)\hat{L}-\hat{L}\dg\F{0}_{z}(t)\Big{]}\lpsizk{t},
\end{equation} where the operator functional $\F{0}_{z}(t)$ is defined as
\begin{eqnarray} \label{F}
\F{0}_{z}(t)=\int_{0}^{t}\alpha(t-s)\f{0}_{z}(t,s)ds.
\end{eqnarray} The significance of the superscripts $(0)$ proceeding these operators will
become apparent in Sec. \ref{Pert}.

To derive the actual (non-linear) non-Markovian SSE we need to condition the state on a
noise function that is equivalent to the actual probability distribution, \begin{equation}
\label{Probability} {P}(\{a_{k}\},t)=\langle{\Psi(t)}\ak\ab{\Psi(t)}\rangle.
\end{equation} For most systems $\ket{\Psi(t)}$ is unknown. Nevertheless we
can use a Girsanov transformation \cite{GamWis02,DioGisStr98} to relate the actual noise
function to the ostensible noise function. In this case,
\begin{equation}\label{Cnoise} z(t)=z_{\Lambda}(t)+\int_{0}^{t} \alpha{(t-s)} \langle \hat{L}\rangle_s
ds ,
\end{equation} where $z_{\Lambda}(t)$ is equivalent to the noise function used in the
ostensible case, satisfying the correlations defined in
\erfs{CCorrelation1}{CCorrelation2}. With the correct $z(t)$ the actual non-Markovian SSE
for the normalised state  is \cite{GamWis02,DioGisStr98}
\begin{eqnarray}\label{CSSEOperatorForm} d_t\psizk{t}&=&\Big{[}
{-\frac{i}{\hbar}\hat{H}_{\rm int}(t)}-
(\hat{L}\dg-\langle\hat{L}\dg\rangle_t) \F{0}_{z}(t) \nl+
\Big{\langle}(\hat{L}\dg-\langle\hat{L}\dg\rangle_t) \F{0}_{z}(t)
\Big{\rangle}_t+z^{*}(t)\nl\times(\hat{L}-\langle\hat{L}\rangle_t)\Big{
]}\psizk{t}.
\end{eqnarray} The notation $\langle
\hat{L}\rangle_t$ is short hand for $\psizb{t}\hat{L}\psizk{t}$. From \erf{CLSSEOperatorForm} and \erf{CSSEOperatorForm}
if the operator functional $\F{0}_{z}(t)$ is known for all time and for each noise function $z(t)$  we can solve the
coherent non-Markovian SSE.

\subsubsection{Quadrature Unravelling-Outlined}

To obtain a non-Markovian SSE with real noise, it is natural to consider a quadrature
noise operator,\begin{equation} \label{QuadratureNoiseOpeator} \hat{z}(t)={\hat{b}_{\rm
int }(t)e^{i\omega_{0} t}e^{-i\phi}+\hat{b}_{\rm int}\dg(t)e^{-i\omega_{0}
t}e^{i\phi}},\end{equation} where $\hat{b}_{\rm int}(t)$ is defined in equation
(\ref{BathOperatorInt}) and $\phi$ is some arbitrary phase. The phase $\phi$ defines the
measured quadrature: an $x$-quadrature measurement occurs when $\phi$ is set to zero, and
the conjugate measurement of the $y$-quadrature occurs when $\phi=\pi/2$. Unless otherwise
stated we will set $\phi$ to zero.

The measurement basis for the bath measurement is $\qk$ and must satisfy \begin{equation}
\label{basis} \hat{z}(t)\qk=z(t)\qk.
\end{equation} The problem with this noise function is that in general it is hard (maybe impossible) to work out a time-independent
eigenstate in the interaction picture. However, we can find this eigenstate if we make the assumptions that for every
mode $k$ there exists another mode, which we can label $-k$, such that $\Omega_{-k}=-\Omega_{k}$ and $g_{-k}=g^{*}_{k}$.
These assumptions simply mean that the modes coupled to the system come in symmetric pairs about the frequency $\Omega$.
Without loss of generality we can take the $g_{k}$'s to be real, absorbing any phases in the definitions of the bath
operators. With all of these assumptions we can rewrite equation (\ref{QuadratureNoiseOpeator}) as
\begin{equation}\label{QuadratureNoiseOperatorXY} \hat{z}(t)=\sum_{k>0} {2} g_{k}[
\hat{X}_{k}^{+} \cos(\Omega_{k}t)+\hat{Y}_{k}^{-}\sin(\Omega_{k}t)]. \end{equation} Here
we have introduced the two-mode quadrature operators \begin{subequations} \begin{eqnarray}
\hat{X}_{k}^{\pm}&=&(\hat{x}_{k}\pm\hat{x}_{-k})/{\rt{2}}\label{X+}, \\
\hat{Y}_{k}^{\pm}&=&(\hat{y}_{k}\pm\hat{y}_{-k})/{\rt{2}}, \label{Y-} \end{eqnarray}
\end{subequations} where $\hat{x}_{k}$ and $\hat{y}_{k}$ are the quadratures of
$\hat{a}_{k}$: \begin{equation} \label{aIintoxAndy} \hat{a}_{k}=({\hat{x}_{k}+i
\hat{y}_{k}})/{\rt{2}}. \end{equation} The measurement basis that satisfies \erf{basis},
in the x-quadrature representation is \begin{equation} \label{Qstatex} \qk=\prod_{k>0}\int
\frac{dx'}{\rt{2\pi}}
\Big{|}\frac{X_{k}^{+}-x'}{\rt{2}}\Big{\rangle}_{-k}\Big{|}\frac{X_{k}^{+}+x'}{\rt{2}
}\Big{\rangle}_{k}e^{iY_{k}^{-}x'}. \end{equation} With this basis and the above noise
operator the noise function for the quadrature measurement is \begin{equation}
\label{QuadratureNoiseFunction} z(t)=\sum_{k>0} {2} g_{k} [  X_{k}^{+}
\cos(\Omega_{k}t)+Y_{k}^{-}\sin(\Omega_{k}t)], \end{equation} which by definition is real.

Furthermore under the above assumptions the memory function $\alpha(t-s)$ in
\erf{MemoryFunction} reduces to
\begin{equation}\label{MemoryFunctionAssump} \beta{(t-s)}=2\sum_{k>0}|g_{k}|^{2}\cos[\Omega_k (t-s)]. \end{equation}
As in the coherent case we define the ostensible distribution as the overlap between the
vacuum state and $\qk$, that is \beq \label{QuadratureProbability}
\Lambda(\{X_{k},Y_{k}\}) = \pi^{-K/2}{e^{-\sum_{k>0} ({X_{k}^{+}}^2+{Y_{k}^{-}}^2)}}. \eeq
With this distribution the correlation for the real-valued noise function is
\begin{equation} \label{NMNoiseCorrelation} \tilde{E}[z(t)z(s)]=\beta{(t-s)}, \end{equation} where the tilde, like
before, means an average over the ostensible distribution. For
this ostensible distribution the differential equation for
$\lpsizk{t}$ is \begin{eqnarray} \label{QLSSEFunctionalForm}
\partial_{t}\lpsizk{t}& =&\Big{[}{-\frac{i}{\hbar}\hat{H}_{\rm
int}(t)}+{z}(t)\hat{L}
-\hat{L}_{x}\int_{0}^{t}\beta{(t-s)}\nl\times \frac{\delta}
{\delta z(s)} ds\Big{]}\lpsizk{t}, \end{eqnarray} where
$\hat{L}_{x}=\hat{L}+\hat{L}\dg$. Making the Ansatz,
\begin{equation}\label{AnsatzQ} \frac{\delta}{\delta
z(s)}\lpsizk{t}=\q{0}_{z}(t,s)\lpsizk{t}, \end{equation} the
linear non-Markovian SSE is \begin{equation}
\label{QLSSEOperatorForm}
\partial_{t}\lpsizk{t} =\Big{[}{-\frac{i}{\hbar}\hat{H}_{\rm int}(t)}+{z}(t)\hat{L}
-\hat{L}_{x}\Q{0}_{z}(t)\Big{]}\lpsizk{t}, \end{equation} where
\begin{eqnarray} \label{Theta} \Q{0}_{z}(t)=\int_{0}^{t}\beta(t-s)\q{0}_{z}(t,s)ds.
\end{eqnarray}

To derive the actual non-Markovian SSE we need to calculate the correct noise function.
The Girsanov transformation
 giving the actual real-valued $z(t)$ is \cite{GamWis02}\begin{eqnarray} \label{TrueNoise} {z}(t)&=&
z_{\Lambda}(t)+\int_{0}^{t}\langle\hat{L}_{x}\rangle_s \beta(t-s) ds, \end{eqnarray} where $z_{\Lambda}(t)$ satisfies
the correlations defined in \erf{NMNoiseCorrelation}. The actual non-Markovian SSE for the quadrature unravelling is
 \begin{eqnarray}\label{QSSEOperatorForm} d_t\psizk{t}&=&\Big{[}
{-\frac{i}{\hbar}\hat{H}_{\rm int}(t)}-
(\hat{L}_{x}-\langle\hat{L}_{x}\rangle_t) \Q{0}_{z}(t) \nl+
\Big{\langle}(\hat{L}_x-\langle\hat{L}_x\rangle_t) \Q{0}_{z}(t)
\Big{\rangle}_t
+z(t)\nl\times(\hat{L}-\langle\hat{L}\rangle_t)\Big{ ]}\psizk{t}.
\end{eqnarray} Thus, if $\Q{0}_{z}(t)$ is known for $z(t)$ and all
time then we can solve the quadrature non-Markovian SSE.

\section{Perturbation Method}\label{Pert}

To solve the non-Markovian SSE, and hence find $\rho_{\rm red}(t)$, for the coherent or quadrature unravelling we have
to work out the operator functionals $\F{0}_{z}(t)$ and $\Q{0}_{z}(t)$ respectively. This has been done exactly only for
systems for which an analytical solution for $\rho_{\rm red}(t)$ may be found by other means
\cite{DioGisStr98,StrDioGis99,Cre00} or for systems with a small number of bath modes \cite{GamWis02}. In this section
we to propose our perturbation technique for working out these functionals when exact solutions are not possible.

\subsection{Perturbation Approach for the Coherent Unravelling}

The perturbation that we are going to propose is only valid for memory functions of the form
\begin{subequations} \label{memoryFunctForm}
\begin{equation}
  \alpha(t-s)=\sum_{j=1}^{J}\alpha^{(j)}(t-s),
\end{equation} where
\begin{equation}\label{partMemoryFunctForm}
\alpha^{(j)}(t-s)=|G_{j}|^{2}e^{-\kappa_{j}|t-s|/2}e^{-i\Omega_{j}(t-s)}.
\end{equation}
\end{subequations}
In principle this is always a valid decomposition for the memory function as in the
$J\rightarrow\infty$ and $\kappa_{j} \rightarrow 0$ limit this memory function approaches
the microscopic memory function displayed in \erf{MemoryFunction}. In Ref. \cite{SteIma96}
the authors suggest that in practice most environments can be simulated with $J$ being
quite
small.

With this expansion for the memory function the functional $\F{0}_{z}(t)$ can be written as
\begin{equation}\label{FinFj}
  \F{0}_{z}(t)=\sum_{j}\F{0}_{z}^{(j)}(t).
\end{equation} where
\begin{eqnarray}  \label{Fj}
  \F{0}_{z}^{(j)}(t)=\int_{0}^{t}\alpha^{(j)}(t-s)\f{0}_{z}(t,s)ds.
\end{eqnarray}
To calculate these operator functionals we set up a set of coupled nonlinear differential
equations for
$\F{0}_{z}^{(j)}(t)$. Taking the time derivative of \erf{Fj} we get
\begin{eqnarray}\label{Fjdot} \partial_t
\F{0}_{z}^{(j)}(t)&=&\alpha^{(j)}(0)\f{0}_{z}(t,t) +\int_{0}^{t}
[\partial_{t}\alpha^{(j)}(t-s)]\nl\times
~^{(0)}\hat{f}_{z}(t,s)ds+\int_{0}^{t} \alpha^{(j)}(t-s)\nl\times
\partial_{t}\f{0}_{z}(t,s)ds.\hspace{.8cm}
\end{eqnarray}
The first term is easily evaluated using \begin{equation}\label{1}
  ^{(0)}\hat{f}_{z}(t,t)=\hat{L},
\end{equation}as derived in Appendix \ref{AA}. The second term is where our earlier
decomposition of $\alpha(t-s)$ is used. We chose $\alpha^{(j)}(t-s)$ such that
$\partial_{t}\alpha^{(j)}(t-s)\propto\alpha^{(j)}(t-s)$. This results in the second term equaling
\begin{equation}\label{2}
-(\frac{\kappa_{j}}{2}+i\Omega_{j})\F{0}_{z}^{(j)}(t).
\end{equation}

The third term involves the partial derivative
$\partial_{t}[\f{0}_{z}(t,s)]$. To find this we use the fact that
\begin{equation} \label{CONCON}
\partial_{t}\frac{\delta}{\delta z^{*}(s)}\lpsizk{t}=\frac{\delta}{\delta
z^{*}(s)}\partial_{t}\lpsizk{t},
\end{equation} which is called the consistency condition in \cite{DioGisStr98}.
 This consistency condition is only valid for $t\neq s$ this
is because at time $t=s$ the functional derivative is not well defined. Using Eq.
(\ref{AnsatzO}) we can write the left-handed side (LHS) of the consistency condition as
\begin{eqnarray}
\partial_{t}\frac{\delta}{\delta z^{*}(s)}\lpsizk{t}
&=&[\partial_t\f{0}_{z}(t,s)]\lpsizk{t}\nl+
\f{0}_{z}(t,s)\partial_{t}\lpsizk{t}.
\end{eqnarray}
Substituting \erf{CLSSEOperatorForm} in for $\partial_{t}\lpsizk{t}$ gives
\begin{eqnarray}
\partial_{t}\frac{\delta}{\delta z^{*}(s)}\lpsizk{t}&=&\Big{[}\partial_t\f{0}_{z}(t,s)
-\frac{i}{\hbar}\f{0}_{z}(t,s)\hat{H}_{\rm int}(t)
\nl+z^{*}(t)\f{0}_{z}(t,s)\hat{L}-
\f{0}_{z}(t,s)\nl\times\hat{L}\dg\F{0}_{z}(t)\Big{]} \lpsizk{t}.
\end{eqnarray}
Using \erfs{CLSSEOperatorForm}{AnsatzO} the right-handed side (RHS) of the consistency condition gives
\begin{eqnarray}
\frac{\delta}{\delta z^{*}(s)}\partial_{t}\lpsizk{t}&=&
\Big{[}-\frac{i}{\hbar}\hat{H}_{\rm int }(t)\f{0}_{z}(t,s) +
z^{*}(t)\hat{L}\nl\times\f{0}_{z}(t,s)
-\hat{L}\dg\F{0}_{z}(t)\f{0}_{z}(t,s)\nl-\hat{L}\dg\frac{\delta}{\delta
z^{*}(s)}\F{0}_{z}(t) \Big{]}\lpsizk{t}.
\end{eqnarray}
Equating the LHS with the RHS gives
\begin{eqnarray} \label{ComCoh}
\partial_{t}\f{0}_{z}(t,s)&=& -\frac{i}{\hbar} [\hat{H}_{\rm
int}(t),\f{0}_{z}(t,s)]+z^{*}(t)[\hat{L},\nl\f{0}_{z}(t,s)]-
[\hat{L}\dg\F{0}_{z}(t),\f{0}_{z}(t,s)]\nl-\hat{L}\dg\frac{\delta}{\delta
z^{*}(s)}\F{0}_{z}(t).
\end{eqnarray}
Substituting this equation with \erfs{1}{2} into \erf{Fjdot} we get
\begin{eqnarray}\label{FjdotFinal}
\partial_t\F{0}^{(j)}_{z}(t)
 &=& |G_{j}|^{2}\hat{L}
 -(\frac{\kappa_{j}}{2}+i\Omega_{j})\F{0}^{(j)}_{z}(t)
+ z^{*}(t)\nl\times[\hat{L},\F{0}^{(j)}_{z}(t)]-\frac{i}{\hbar}
[\hat{H}_{\rm
int}(t),\F{0}^{(j)}_{z}(t)]\nl-[\hat{L}\dg\F{0}_{z}(t),\F{0}^{(j)}_{z}(t)]
-\hat{L}\dg\nl\times\sum_{k}\F{1}_{z}^{(j,k)}(t),\nl
\end{eqnarray}
where $\F{1}_{z}^{(j,k)}(t)$ is our first order functional. It has the form
\begin{eqnarray} \label{Fj1}
  \F{1}_{z}^{(j,k)}(t)
  =\int_{0}^{t}\alpha^{(j)}(t-s)\f{1}^{(k)}_{z}(t,s) ds,
\end{eqnarray} where we have used the following Ansatz
\begin{equation}\label{Ansatz1}
  \frac{\delta}{\delta
z^{*}(s)}\F{0}^{(k)}_{z}(t)=\f{1}^{(k)}_{z}(t,s).
\end{equation}
If we knew the form of $\F{1}_{z}^{(j,k)}(t)$ then \erf{FjdotFinal} could be solved numerically.

To find the form of $\F{1}_{z}^{(j,k)}(t)$ we can take the time
 derivative of \erf{Fj1}. Doing this we get
\begin{eqnarray}\label{1Fjndot} \partial_t
 \F{1}_{z}^{(j,k)}(t)&=&\alpha^{(j)}(0)\f{1}^{(k)}_{z}(t,t)
 +\int_{0}^{t} [\partial_{t}\alpha^{(j)}(t-s)]\nl\times \f{1}^{(k)}_{z}(t,s)ds
+\int_{0}^{t} \alpha^{(j)}(t-s)\nl\times
\partial_{t} \f{1}^{(k)}_{z}(t,s)ds.\hspace{.5cm}
\end{eqnarray}
The first term is easy to work out. From \erf{FjdotFinal} it follows that
\begin{equation}
\f{1}^{(k)}_{z}(t,t)=[\hat{L},\F{0}^{(k)}_{z}(t)].
\end{equation} The second term as before also simply evaluates to
\begin{equation}
  -\Big{(}\frac{\kappa_{j}}{2}+i\Omega_{j}\Big{)}\F{1}^{(j,k)}_{z}(t).
\end{equation}
The third term is worked out via a new consistency condition,
\begin{equation} \label{CONCON1}
\partial_{t}\frac{\delta}{\delta z^{*}(s)}\F{0}^{(k)}_{z}(t)=\frac{\delta}{\delta
z^{*}(s)}\partial_{t}\F{0}^{(k)}_{z}(t).
\end{equation}
Substituting \erfs{Ansatz1}{FjdotFinal} into this consistency
condition gives \begin{widetext}\begin{eqnarray}
\partial_{t}\f{1}^{(k)}_{z}(t,s)
 &=&-(\frac{\kappa_{k}}{2}+i\Omega_{k})\f{1}_{z}^{(k)}(t,s)
-\frac{i}{\hbar} [\hat{H}_{\rm
int}(t),\f{1}^{(k)}_{z}(t,s)]+z^{*}(t)[\hat{L},\f{1}^{(k)}_{z}(t,s)]\nl-
[\hat{L}\dg\sum_{l}\f{1}^{(l)}_{z}(t,s),\F{0}^{(k)}_{z}(t)] -
[\hat{L}\dg\sum_{l}\F{0}^{(l)}_{z}(t),\f{1}^{(k)}_{z}(t,s)]\nl
-\hat{L}\dg\sum_{l}\frac{\delta}{\delta
z^{*}(s)}\F{1}_{z}^{(k,l)}(t).\hspace{.5cm}
\end{eqnarray}
Substituting all these terms into \erf{1Fjndot} gives \begin{eqnarray} \label{1Fjdotfinal}
\partial_t
 \F{1}_{z}^{(j,k)}(t)&=&|G_{j}|^2[\hat{L},\F{0}^{(k)}_{z}(t)]
 -\Big{(}\frac{\kappa_{j}}{2}+i\Omega_{j}\Big{)}\F{1}^{(j,k)}_{z}(t)
-\Big{(}\frac{\kappa_{k}}{2}+i\Omega_{k}\Big{)}\F{1}_{z}^{(j,k)}(t)\nl
-\frac{i}{\hbar} [\hat{H}_{\rm
int}(t),\F{1}^{(j,k)}_{z}(t)]+z^{*}(t)[\hat{L},\F{1}^{(j,k)}_{z}(t)]-
[\hat{L}\dg\sum_{l}\F{1}^{(j,l)}_{z}(t),\F{0}^{(k)}_{z}(t)] \nl-
[\hat{L}\dg\sum_{l}\F{0}^{(l)}_{z}(t),\F{1}^{(j,k)}_{z}(t,s)]
-\hat{L}\dg\sum_{l}\F{2}_{z}^{(j,k,l)}(t).
\end{eqnarray}\end{widetext}
Where the last term is the second order functional, which equals
\begin{equation} \label{Fj2}
  \F{2}_{z}^{(j,k,l)}(t)
  =\int_{0}^{t}\alpha^{(j)}(t-s)\frac{\delta}{\delta
z^{*}(s)} \F{1}^{(k,l)}_{z}(t)ds.
\end{equation}

Here we see that we can develop a general way for setting up an $n^{\rm th}$ order differential equations. The $n^{\rm
th}$ order functional is
\begin{equation} \label{Fjn}
  \F{n}_{z}^{(j,k,...,l)}(t)
  =\int_{0}^{t}\alpha^{(j)}(t-s)\f{n}^{(k,...,l)}_{z}(t,s) ds,
\end{equation} where we have used the Ansatz
\begin{equation}\label{Ansatzn}
  \frac{\delta}{\delta
z^{*}(s)}\F{n-1}^{(k,...,l)}_{z}(t)=\f{n}^{(k,...,l)}_{z}(t,s).
\end{equation}
The differential equation for the $n^{\rm th}$ order functional is
\begin{eqnarray}\label{nFjndot} \partial_t
 \F{n}_{z}^{(j,k,...,l)}(t)&=&\alpha^{(j)}(0)\f{n}^{(k,...,l)}_{z}(t,t)\nl\hspace{-1cm}
 +\int_{0}^{t}[ \partial_{t}\alpha^{(j)}(t-s)] \f{n}^{(k,...,l)}_{z}(t,s)ds
 \nl\hspace{-1cm}+\int_{0}^{t} \alpha^{(j)}(t-s)
\partial_{t}\f{n}^{(k,...,l)}_{z}(t,s)ds.\nl
\end{eqnarray}
The first term can always be calculated by the $(n-1)^{\rm th}$ differential equation. The second term is always simple
to calculate as $\partial_{t}\alpha^{(j)}(t-s)\propto\alpha^{(j)}(t-s)$ and the third term is always calculable by
 the $(n-1)^{\rm th}$ order consistency condition
  \begin{equation}
\partial_{t}\frac{\delta}{\delta z^{*}(s)}\F{n-1}^{(k,...,l)}_{z}(t)=\frac{\delta}{\delta
z^{*}(s)}\partial_{t}\F{n-1}^{(k,...,l)}_{z}(t).
\end{equation}

The $n^{\rm th}$ order perturbation method propose is to terminate this series by setting $\F{n}_{z}^{(j,k,...,l)}(t)$
equal to an arbitrary operator. The simplest scheme would be to set this operator to zero, but to keep the theory
consistent with the Markov limit for all orders, we set $\F{n}_{z}^{(j,k,...,l)}(t)$ in the following manner. The zeroth
order perturbation aries when we use the approximation
\begin{equation}\label{pertbat0}
\F{0}_{z}^{(j)}(t)
\simeq\int_{0}^{t}\alpha^{(j)}(t-s)ds\frac{\delta }{\delta
z^{*}(t)}=\int_{0}^{t}\alpha^{(j)}(t-s)ds\hat{L}.
\end{equation} Note that the approximation here is the replacement of $\delta/\delta z^{*}(s)$ by $\delta/\delta
z^{*}(t)$. The first order perturbation arises when we use the approximation
\begin{eqnarray}\label{pertbat1}
\F{1}_{z}^{(j,k)}(t)&\simeq&\int_{0}^{t}\alpha^{(j)}(t-s)ds\frac{\delta
\F{0}^{(k)}_{z}(t)}{\delta z^{*}(t)}\nn\\
&=&\int_{0}^{t}\alpha^{(j)}(t-s)ds[\hat{L},\F{0}^{(k)}_{z}(t)]
\end{eqnarray} and $\F{0}_{z}^{(j)}(t)$ is calculated via \erf{FjdotFinal}.
The $n^{\rm th}$ order perturbation arises when we use the approximation
\begin{eqnarray}\label{pertbatn}
\F{n}_{z}^{(j,k,...,l)}(t)
&\simeq&\int_{0}^{t}\alpha^{(j)}(t-s)ds\frac{\delta
\F{n-1}^{(k,...,l)}_{z}(t)}{\delta z^{*}(t)}\nn\\
&=&\int_{0}^{t}\alpha^{(j)}(t-s)ds[\hat{L},\F{n-1}^{(k,...,l)}_{z}(t)]\nl
\end{eqnarray} and $\F{0}_{z}^{(j)}(t),...,\F{n-1}_{z}^{(j,...,k)}(t)$ are calculated via Eqs.
(\ref{FjdotFinal}), (\ref{1Fjdotfinal}) and (\ref{nFjndot}).
 The physical motivations for choosing this type of expansion are;
 \\ $~~~~$ a) For most system the memory function will decay and thus the most dominant term in the functional derivative
will be the
value as $s\rightarrow t$.\\
 $~~~~$b) Only $\F{0}^{(j)}_{z}(t)$ affects the system directly,
 so the further removed the approximation the more accurate we
 expect the approximation to be.\\
 $~~~~$c) In the Markovian limit, only the zero order term is needed.

To summarize this perturbation method, for environments which can
be modelled by \erf{memoryFunctForm}, it is possible to obtain a
perturbative solution for the coherent non-Markovian SSE. From
these SSEs it is possible to generate a perturbative solution for
$\rho_{\rm red}(t)$, which by definition will always be positive.
The number of coupled complex differential equations that are
required for this technique is
\begin{equation}
  d^{2}(J^{n}+J^{n-1}+...+J)+d+J=d^{2}J\frac{J^{n}-1}{J-1}+d+J
\end{equation}  where $d$ is the system
dimension, $J$ is the number of exponentials required to simulate
the memory function and $n$ is the order of the perturbation. The
first term represents the number of equations needed to simulate
the functional derivative. The next term $d$ is for the $d$
complex amplitudes of the system. The final term $J$ is for the
stochastic equations needed to generate the noise function $z(t)$.

\subsection{Perturbation Approach for the Quadrature Unravelling}

The perturbation method in the quadrature case is essentiality the same as the coherent
case, but the memory function expressed in \erf{partMemoryFunctForm} is too general. This
is because the memory function for the quadrature unraveling must be consistent with the
assumptions stated below \erf{basis}. The most general memory function that satisfies
these requirements is
\begin{subequations} \label{BmemoryFunctForm}
\begin{equation}
  \beta(t-s)=\sum_{j}^{'}\beta^{(j,\cos)}(t-s),
\end{equation} where
\begin{equation}\label{BpartMemoryFunctForm}
\beta^{(j,\cos)}(t-s)=2|G_{j}|^{2}e^{-\kappa_{j}|t-s|/2}\cos[\Omega_{j}(t-s)].
\end{equation}
\end{subequations}
This presents a problem as $\partial_{t}\beta^{(j,\cos)}(t-s)$ is not proportional to
$\beta^{(j,\cos)}(t-s)$. To get around this we define a new function
$\beta^{(j,\sin)}(t-s)$ as
\begin{equation}\label{BSinFunctForm}
\beta^{(j,\sin)}(t-s)=2|G_{j}|^{2}e^{-\kappa_{j}|t-s|/2}\sin(\Omega_{j}(t-s)),
\end{equation} and two functionals
\begin{subequations}
\begin{eqnarray}
\label{Thetaj} \Q{0}_{z}^{(j,\cos)}(t)&=&\int_{0}^{t}\beta^{(j,\cos)}(t-s)\hat{q}_{z}(t,s)ds, \\
\label{thetaS}
\Q{0}_{z}^{(j,\sin)}(t)&=&\int_{0}^{t}\beta^{(j,\sin)}(t-s)\hat{q}_{z}(t,s)ds.
\end{eqnarray}
\end{subequations}
The functional $\Q{0}_{z}(t)$ is the found by
\begin{equation}\label{QinQj}
  \Q{0}_{z}(t)=\sum_{j}\Q{0}_{z}^{(j,\cos)}(t).
\end{equation}
Taking the time derivative of \erfs{Thetaj}{thetaS} we get
\begin{subequations}\label{Qjdot}
\begin{eqnarray}\label{Qjjdot}
  d_t \Q{0}^{(j,\cos)}_{z}(t)&=&\beta^{(j,\cos)}(t,t)\q{0}_{z}(t,t)
  \nl+\int_{0}^{t} [\partial_{t}\beta^{(j,\cos)}(t-s)]
  \q{0}_{z}(t,s)ds\nl+\int_{0}^{t} \beta^{(j,\cos)}(t-s)
  \partial_{t}\q{0}_{z}(t,s)ds,\nl \\ \label{QSdot}
  d_t \Q{0}^{(j,\sin)}_{z}(t)&=&\int_{0}^{t} [\partial_{t}\beta^{(j,\sin)}(t-s)]
 \q{0}_{z}(t,s)ds\nl+\int_{0}^{t} \beta^{(j,\sin)}(t-s)
  \partial_{t}\q{0}_{z}(t,s)ds.\nl
\end{eqnarray}
\end{subequations}
As in the coherent case it can be shown that $\q{0}_{z}(t,t) =\hat{L}$. The two terms
involving the derivative of $\beta^{(j,\cos)}(t-s)$ and $\beta^{(j,\sin)}(t-s)$ by
definition give
\begin{subequations}
\begin{eqnarray}
  \int_{0}^{t} \partial_{t}\beta^{(j,\cos)}(t-s)
  \q{0}_{z}(t,s)ds&=&
  -\frac{\kappa_{j}}{2}\Q{0}^{(j,\cos)}_{z}(t)\nl\hspace{-2cm}
  -\Omega_{j}\Q{0}^{(j,\sin)}_{z}(t) \\
  \int_{0}^{t} \partial_{t}\beta^{(j,\sin)}(t-s)
  \q{0}_{z}(t,s)ds&=&-\frac{\kappa_{j}}{2}\Q{0}^{(j,\sin)}_{z}(t)\nl\hspace{-2cm}
  +\Omega_{j}\Q{0}^{(j,\cos)}_{z}(t).\hspace{.8cm}
\end{eqnarray}
\end{subequations} The last two terms require finding $\partial_{t}\q{0}_{z}(t,s)$. As
in the coherent
case this is found by the consistency condition
\begin{equation}
\partial_{t}\frac{\delta}{\delta z(s)}\lpsizk{t}=\frac{\delta}{\delta
z(s)}\partial_{t}\lpsizk{t},
\end{equation} yielding
\begin{eqnarray} \label{ComQuad}
 \partial_{t}\q{0}_{z}(t,s)&=&
-\frac{i}{\hbar} [\hat{H}_{\rm
int}(t),\q{0}_{z}(t,s)]+z(t)[\hat{L},\nl\q{0}_{z}(t,s)]-
[\hat{L}_{x}\Q{0}_{z}(t),\q{0}_{z}(t,s)]\nl-\hat{L}_{x}\frac{\delta}{\delta
z(s)}\Q{0}_{z}(t).
\end{eqnarray}
Substituting these terms into \erf{Qjdot} we get
\begin{widetext}\begin{subequations}\label{QjdotFinal}
\begin{eqnarray}
d_t \Q{0}^{(j,\cos)}_{z}(t)&=&2|G_{j}|^2\hat{L}
   -\frac{\kappa_{j}}{2}\Q{0}^{(j,\cos)}_{z}(t)
  -\Omega_{j}\Q{0}^{(j,\sin)}_{z}(t)
-\frac{i}{\hbar} [\hat{H}_{\rm
int}(t),\Q{0}^{(j,\cos)}_{z}(t)]\nl+z(t)[\hat{L},\Q{0}^{(j,\cos)}_{z}(t)]-
[\hat{L}_{x}\Q{0}_{z}(t),\Q{0}^{(j,\cos)}_{z}(t)]
\nl-\hat{L}_{x}\sum_{k}\Q{1}^{(j,k,\cos,\cos)}_{z}(t)
 ,
\\
  d_t \Q{0}^{(j,\sin)}_{z}(t)&=&-\frac{\kappa_{j}}{2}\Q{0}^{(j,\sin)}_{z}(t)
  +\Omega_{j}\Q{0}^{(j,\cos)}_{z}(t)
  -\frac{i}{\hbar} [\hat{H}_{\rm
int}(t),\Q{0}^{(j,\sin)}_{z}(t)]+z(t)[\hat{L},\Q{0}^{(j,\sin)}_{z}(t)]\nl-
[\hat{L}_{x}\Q{0}_{z}(t),^{(0)}\Q{0}^{(j,\sin)}_{z}(t)]
-\hat{L}_{x}\sum_{k}~\Q{1}^{(j,k,\sin,\cos)}_{z}(t),
\end{eqnarray}
\end{subequations} \end{widetext}
where
\begin{subequations}
\begin{equation}
\Q{1}^{(j,k,\cos,\cos)}_{z}(t)=\int_{0}^{t}\beta^{(j,\cos)}(t,s)
\frac{\delta\Q{0}^{(k,\cos)}_{z}(t)}{\delta z(s)}ds,
\end{equation}
\begin{equation}
\Q{1}^{(j,k,\sin,\cos)}_{z}(t)=\int_{0}^{t}\beta^{(j,\sin)}(t,s)\frac{\delta\Q{0}^{(k,\cos)}_{z}(t)}{\delta
z(s)}ds.
\end{equation}\end{subequations}
The higher order functional differential equations are found in the same manner as in the
coherent case, except the form of $\beta(t-s)$ results in $2^{n}$ as many equations for
order $n$.

The perturbation expansion is similar for this unravelling, the only difference being that
we have $2^{n}$ operators to approximate. The $0^{\rm th}$ order approximation is to set
the $0^{\rm th}$ order functionals to
\begin{subequations}\begin{eqnarray}
\Q{0}^{(j,\cos)}_{z}(t)&=&\int_{0}^{t}\beta^{(j,\cos)}(t,s)ds\hat{L}\\
\Q{0}^{(j,\sin)}_{z}(t)&=&\int_{0}^{t}\beta^{(j,\sin)}(t,s)ds\hat{L}.
\end{eqnarray}\end{subequations} The first order approximation is to
set the four first order functionals to
\begin{subequations}\begin{eqnarray} \Q{1}^{(j,k,\cos,\cos)}_{z}(t)
&=&\int_{0}^{t}\beta^{(j,\cos)}(t,s)ds[\hat{L},\Q{0}^{(k,\cos)}_{z}(t)],\nl\\
\Q{1}^{(j,k,\sin,\cos)}_{z}(t)&=&\int_{0}^{t}\beta^{(j,\sin)}(t,s)ds[\hat{L},\Q{0}^{(k,\cos)}_{z}(t)],\nl\\
\Q{1}^{(j,k,\cos,\sin)}_{z}(t)&=&\int_{0}^{t}\beta^{(j,\cos)}(t,s)ds[\hat{L},\Q{0}^{(k,\sin)}_{z}(t)],\nl\\
\Q{1}^{(j,k,\sin,\sin)}_{z}(t)&=&\int_{0}^{t}\beta^{(j,\sin)}(t,s)ds
[\hat{L},\Q{0}^{(k,\sin)}_{z}(t)].\nl
\end{eqnarray}\end{subequations}
and we calculate the ${0^{\rm th}}$ order functionals via \erf{QjdotFinal}.

\section{Enlarged System Approach} \label{Enl}

To test the accuracy of our perturbation method we compare our
results for the reduced state with the reduced state found via the
enlarged system method of Imamo$\bar{\rm g}$lu
\cite{Ima94,SteIma96}. An example of how this method is applied to
a non-Markovian system can be found in Ref. \cite{CaoLonWeiCao01}.

For those who are not familiar with the enlarged system method, we provide a short proof
that the reduced system dynamics are exactly reproduced by the enlarged system method
provided that $\alpha(t-s)$, called $\Gamma(\tau)$ in Refs \cite{Ima94,SteIma96}, is of
the form
\begin{equation}\label{memoryFunctForm2}
\alpha(t-s)=\sum_{j}|G_{j}|^2 e^{-\kappa_{j}|t-s|/2-i\Omega_{j}(t-s)},
\end{equation} which is the same as \erf{memoryFunctForm}.

The total Hamiltonian for the enlarged system is
\begin{eqnarray}\label{enlham}
  \hat{H}_{\rm tot}&=&
  \hat{H}_{\rm sys}+\hbar\sum_{j}\omega_{j}\hat{c}_{j}\dg\hat{c}_{j}
  +\hbar\sum_{j}\int_{-\infty}^{\infty}d\omega\nl\times\omega\hat{\nu}_{j}(\omega)\dg\hat{\nu}_{j}(\omega)
 +i\hbar\sum_{j}[G^{*}_{j}\hat{L}\hat{c}_{j}\dg-G_{j}\hat{L}\dg\hat{c}_{j}]\nl
 +i\hbar\sum_{j}\int_{-\infty}^{\infty}d\omega \rt{\frac{\kappa_{j}}{2\pi}}
  [\hat{\nu}\dg_{j}(\omega)\hat{c}_{j}-\hat{\nu}_{j}(\omega)\hat{c}_{j}\dg],\nl
\end{eqnarray} where $ \hat{H}_{\rm sys}=\hat{H}_{\Omega}+\hat{H}$, $\hat{c}_{j}$
is the annihilation operator for the $j^{\rm th}$ added oscillator and
$\hat{\nu}_{j}(\omega)$ is the Markovian bath operator with the correlation
\begin{equation}
  [\hat{\nu}_{j}(\omega),\hat{\nu}_{k}\dg(\omega)]=\delta_{j,k}\delta(\omega-\omega').
\end{equation}
If this is to be the same as \erf{HamiltonianTotal}, then the first two lines of
\erf{enlham} must give $\hat{H}_{\rm sys}+\hat{H}_{\rm bath}$ and the final line
$\hat{V}$. Going to the same interaction picture as we did in Sec. \ref{dynamics}, that is
with respect to the Hamiltonians $\hat{H}_{\Omega}$ and $\hat{H}_{\rm bath}$, we get
\begin{equation}
\hat{V}_{\rm int}(t)=
  i\hbar\sum_{j}[G^{*}_{j}\hat{L}e^{-i\Omega t}\hat{c}_{j}(t)\dg-G_{j}\hat{L}\dg
  e^{i\Omega t}\hat{c}_{j}(t)].
\end{equation} Comparing with \erf{HamiltonianInteractionInt}, for the
enlarged system method to be correct we need $\hat{b}_{\rm
int}(t)=\sum_{j}G_{j}\hat{c}_{j}(t)$.
 To calculate $\hat{c}_{j}(t)$ we use the fact that
\begin{equation}\label{cdot}
 d_{t} \hat{c}_{j}(t)=-i\omega_{j}\hat{c}_{j}(t)-\frac{\kappa_{j}}{2}
 \hat{c}_{j}(t)-\rt{\kappa_{j}}\hat{\nu}_{{\rm in},j}(t)
\end{equation} where $\hat{\nu}_{{\rm in},j}(t)$ is the input field which has a
time commutator $[\hat{\nu}_{{\rm in},j}(t),\hat{\nu}\dg_{{\rm
in},k}(s)]=\delta_{j,k}\delta(t-s)$. For a derivation of equation \erf{cdot} see Ref.
\cite{GarCol85}. This can be integrated to give
\begin{eqnarray}
  \hat{c}_{j}(t)&=&\rt{\kappa_{j}}\int_{0}^{t}
  e^{-\kappa_{j}(t-s)/2-i\omega_{j}(t-s)}\hat{\nu}_{{\rm in},j}(s)ds\nl+\hat{c}_{j}
  (0)e^{-\kappa_{j}t/2-i\omega_{j}t}.
\end{eqnarray} It not obvious that $\sum_{j}G_{j}\hat{c}_{j} (t)$ is
 the same as \erf{BathOperatorInt}. However the time commutator for
 the bath operators is
\begin{equation}
  [\hat{b}_{\rm int}(t),\hat{b}\dg_{\rm int}(s)]e^{i\Omega(t-s)}=\alpha(t-s).
\end{equation}
 In terms of the enlarged system this means
 \begin{eqnarray} &&\hspace{-0.5cm}\sum_{j,k} G_{j}G^{*}_{k}[\hat{c}_{j}(t)
,\hat{c}\dg_{k}(s)]e^{i\Omega(t-s)}\nl= \sum_{j} |G_{j}|^{2}
e^{-\kappa_{j}(t+s)/2-i(\omega_{j}-\Omega)(t-s)}[1\nl+
{\kappa_{j}}\int_{0}^{t}\int_{0}^{s}e^{+\kappa_{j}(t'+s')/2+i\omega_{j}(t'-s')}
\delta(t'-s')dt'ds']\nl= \sum_{j}
|G_{j}|^{2}e^{-\kappa_{j}|t-s|/2-i(\omega_{j}-\Omega)(t-s)}\nl=\alpha(t-s),
\end{eqnarray} provide $\alpha(t-s)$ has the form depicted in
\erf{memoryFunctForm2}. It should noted that this result is exact.
It is not necessary to discard initial transients as in the
derivation in Ref \cite{SteIma96}.

Since we have shown that the total Hamiltonian for the enlarged system is equivalent to
the standard non-Markovian, then the total states $\ket{\Psi_{Sch}(t)}$ must be the same.
We can define a reduced state (in the \sch picture) for the enlarged system as $W_{\rm
Sch}(t)$ which has the Markovian master equation
\begin{eqnarray} \label{schMaster}
 d_{t}W_{\rm Sch}(t)&=&-\frac{i}{\hbar}[\hat{H}_{\Omega}+\hat{H}
 +\hbar\sum_{j}\omega_{j}\hat{c}_{j}\dg\hat{c}_{j}+
 i\hbar\sum_{j}\nl\times(G^{*}_{j}\hat{L}\hat{c}_{j}\dg-G_{j}\hat{L}\dg\hat{c}_{j}),W_{\rm
 Sch}(t)]\nl+\sum_{j}\kappa_{j}{\cal D}[\hat{c}_{j}]W_{\rm
 Sch}(t).
\end{eqnarray}
The reduced state for the system in the $\Omega$-interaction
picture is
\begin{equation}
  \rho_{\rm red}(t)=e^{\frac{i}{\hbar}\hat{H}_{\Omega}t}
  Tr_{\rm enl}[W_{\rm Sch}(t)]e^{-\frac{i}{\hbar}\hat{H}_{\Omega}t}= Tr_{\rm enl}[W_{\rm red}(t)],
\end{equation}
where the trace is performed over the added oscillators and
\begin{equation}
  W_{\rm red}(t)=e^{{i}\sum_{j}\omega_{j}\hat{c}_{j}\dg\hat{c}_{j}t+\frac{i}{\hbar}\hat{H}_{\Omega}t}
W_{\rm
Sch}(t)e^{-{i}\sum_{j}\omega_{j}\hat{c}_{j}\dg\hat{c}_{j}t-\frac{i}{\hbar}\hat{H}_{\Omega}t}.
\end{equation}  This allows us to define a new master equation for the reduced state $W_{\rm red}(t)$ as
\begin{eqnarray}
 d_{t}W_{\rm red}(t)&=&[-\frac{i}{\hbar}\hat{H}_{\rm int}
 +\sum_{j}[G^{*}_{j}\hat{L}\hat{c}_{j}\dg e^{i(\omega_{j}-\Omega)t}\nl-
 G_{j}\hat{L}\dg\hat{c}_{j} e^{-i(\omega_{j}-\Omega)t}],W_{\rm
 red}(t)]\nl+\sum_{j}\kappa_{j}{\cal D}[\hat{c}_{j}]W_{\rm
 red}(t).\nl
\end{eqnarray} which can be solved by standard Markovian techniques,
for example quantum trajectories
\cite{DalCasMol92,GarParZol92,Car93}.

\section{Numerical Example: The Driven Two Level Atom} \label{TLAapply}

In this section we apply our theory to a driven TLA with a simple non-Markovian memory
function.
\begin{equation}\label{simplememfunc} \alpha(t-s)=\frac{\gamma \kappa}{4}
e^{i(\omega_{\rm env}-\Omega)(t-s)}e^{-\kappa|t-s|/2},
\end{equation}
where $\omega_{\rm env}$ is the central frequency of the environment, $\kappa$ represent
the exponential decay of bath memory and $\gamma$ is the Markovian limit decay rate. That
is, in the $\kappa\rightarrow\infty$ limit, $\alpha(t-s)=\gamma\delta(t-s)$, which is the
Markovian limit of the memory function \cite{GamWis02}. We choose an interaction picture
such the $\Omega=\omega_{\rm env}$ so that this memory function is simplifies to
\begin{equation}\label{simplememfunc2}
\alpha(t-s)=\frac{\gamma \kappa}{4}e^{-\kappa|t-s|/2},
\end{equation} which is consistent with the quadrature unravelings assumptions. This results in
$\alpha(t-s)=\beta(t-s)$. However before we apply our theory to the TLA let us revise the
standard TLA model.

\subsection{The TLA}

The TLA is one of the most simple quantum systems to envisage. It consists of two levels,
an excited state $\ket{e}$ of energy $\hbar\omega_{e}$ and a ground state $\ket{g}$ of
energy $\hbar\omega_{g}$. We define the difference in these energies as $\hbar\omega_{0}$
and the zero point energy is taken to be the mid point energy
$\hbar(\omega_{e}+\omega_{g})/2=0$. This allows us to define a system Hamiltonian as
\begin{equation}
\hat{H}_{\rm sys}=\hbar\frac{\omega_{0}}{2}\hat{\sigma}_{z}
\end{equation}
where $\hat{\sigma}_{z}=\ket{e}\bra{e}-\ket{g}\bra{g}$ is one of
the spin matrices for the TLA.

Since we are dealing with open quantum systems we consider the
dynamics of the TLA immersed in the electromagnetic field (the
bath). In the \sch picture with the dipole and rotating wave
approximation (RWA) approximation the interaction Hamiltonian is
\begin{eqnarray}
  \hat{V}=i\hbar\sum_{k}
 ( g^{*}_{k}\hat{\sigma}\hat{a}_{k}\dg
 -g_{k}\hat{\sigma}\dg\hat{a}_{k}),
\end{eqnarray} where $\hat{\sigma}$ is the lowering operator for
the TLA. This is the same form as Eq.
(\ref{HamiltonianInteraction}) with $\hat{L}=\hat{\sigma}$, so the
above non-Markovian SSE theory is applicable to this system.

If we have a TLA driven by a classical electromagnetic field the system Hamiltonian for
the TLA under the RWA approximation is
\begin{equation}
  \hat{H}_{\rm sys}=\hbar\frac{\omega_{0}}{2}\hat{\sigma}_{z}
  +\hbar\frac{\chi}{2}[\hat{\sigma}e^{i\omega_{\rm dr}t}
  +\hat{\sigma}\dg e^{-i\omega_{\rm dr}t}],
\end{equation}
where $\chi$ is the Rabi frequency and $\omega_{\rm dr}$ is the
driving frequency of the classical field.  However as shown in Eq.
(\ref{HamiltonianTotal}) we can also write $\hat{H}_{\rm sys}$ as
$\hat{H}_{\Omega}+\hat{H}(t)$. If
$\hat{H}_{\Omega}={\Omega}\hat{\sigma}_{z}/{2}$, then in the
$\Omega$ interaction picture gives
\begin{equation}
 \hat{H}_{\rm int}(t)=\hbar\frac{\omega_{0}-\Omega}{2}\hat{\sigma}_{z}+\hbar\frac{\chi}{2}
 [\hat{\sigma}e^{i(\omega_{\rm dr}t-\Omega t)}
+\hat{\sigma}\dg e^{-i(\omega_{\rm dr}t-\Omega t)}],
\end{equation}
For our purposes we assume $\Omega=\omega_{\rm dr}$. So
\begin{equation}\label{HAMDRIVEN}
 \hat{H}_{\rm int}(t)=\hbar\frac{\Delta}{2}\hat{\sigma}_{z}+\hbar\frac{\chi}{2}
 \hat{\sigma}_{x},
\end{equation} where $\Delta=\omega_{0}-\Omega$ is the detuning.

For the TLA the reduced state can be written in terms of the real
Bloch vector components $x(t)$, $y(t)$, $z(t)$ as
\begin{equation}
  \rho_{\rm red}(t)=\smallfrac{1}{2}[\hat{I}+x(t)\hat{\sigma}_{x}+y(t)\hat{\sigma}_{y}+z(t)\hat{\sigma}_{z}].
\end{equation}

\subsection{Enlarged System Method}

\begin{figure}
\includegraphics[width= .45\textwidth]{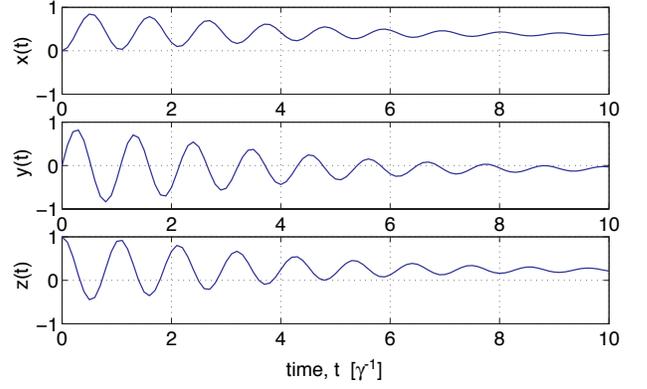}
\caption{\label{Enlarged} This figure depicts the Bloch vector
components of the reduced state of a driven TLA calculated by the
enlarged system method.  In this figure all calculations were done
using the initial system state $\ket{\psi(0)}=\ket{e}$ with system
parameters $\gamma=1$, $\kappa=1$, $\chi=5$ and $\Delta=3$. Time
is measured in units $\gamma^{-1}$.}
\end{figure}

 For the driven TLA with a memory function given by \erf{simplememfunc} the master
equation for the enlarged systems is
\begin{eqnarray}
 d_{t}W_{\rm red}(t)&=&[-\frac{i\Delta}{2}\hat{\sigma}_{z}
 -\frac{i\chi}{2}\hat{\sigma}_{x}
 +\frac{\gamma\kappa}{4}(\hat{\sigma}\hat{c}
 -\hat{\sigma}\dg\hat{c}),\nl W_{\rm
 red}(t)]+\kappa{\cal D}[\hat{c}]W_{\rm
 red}(t).
\end{eqnarray}
 Using $\gamma=1$, $\kappa=1$, $\chi=5$ and $\Delta=3$ the
reduced state is shown in Fig. \ref{Enlarged}. For this simple
case it was noted that the truncation error involved in the
enlarged system state method was negligible. Because of this we
use this reduced state for comparison with the ensemble average of
the non-Markovian SSEs.

\subsection{Coherent Unravelling-TLA}

Applying the coherent non-Markovian SSE theory to the driven TLA, we find that we can
rewrite the actual non-Markovian SSE as
\begin{eqnarray}\label{CSSEOperatorFormTLA}
d_t\psizk{t}&=&\Big{[}
-i\frac{\Delta}{2}\hat{\sigma}_{z}-i\frac{\chi}{2}
 \hat{\sigma}_{x}- (\hat{\sigma}\dg-\langle\hat{\sigma}\dg\rangle_t)
\nl\times\F{0}_{z}(t) +
\Big{\langle}(\hat{\sigma}\dg-\langle\hat{\sigma}\dg\rangle_t)
 \F{0}_{z}(t) \Big{\rangle}_t \nl+z^{*}(t)(\hat{\sigma}-\langle\hat{\sigma}\rangle_t)\Big{ ]}\psizk{t},
\end{eqnarray} and the noise function for the TLA becomes
\begin{equation}\label{CnoiseTLA}
z(t)=z_{\Lambda}(t)+\int_{0}^{t} \alpha{(t-s)} \langle \hat{\sigma}\rangle_s  ds.
\end{equation}

To calculate the complex amplitudes for the actual non-Markovian SSE we apply the system
state $\psizk{t}=C_{e}(t)\ket{e}+C_g(t)\ket{g}$ to \erf{CSSEOperatorFormTLA} and expand
$\F{0}_{z}(t)$ as
\begin{equation}\label{Fexpand}
  \F{0}_{z}(t)=\sum_{m}\hat{m}\Fu{0}_{m,z}(t)
\end{equation} where $m=\{\sigma$, $\sigma\dg$, $\sigma_{z}$, $I\}$. This
results in\begin{widetext}
\begin{subequations} \label{CoherentAmplitudes}
\begin{eqnarray}
d_{t}C_{g}&=&i\frac{\Delta}{2}
C_{g}-i\frac{\chi}{2}C_{e}+z^{*}C_{e}|C_{e}|^2 -
\Fu{0}_{\sigma\dg,z}C^{3}_{g}{C^{*}_{e}}^2+
\Fu{0}_{\sigma,z}C_{g}|C_{e}|^2(1+|C_{e}|^2)\nl-\Fu{0}_{\sigma_{z},z}
C^{2}_{g}C^{*}_{e} (1+2|C_{e}|^2)+\Fu{0}_{I,z}C^{2}_{g}C^{*}_{e},
\hspace{2.0mm}\\
d_{t}C_{e}&=&-i\frac{\Delta}{2} C_{e}-i\frac{\chi}{2}C_{g}
-z^{*}C^{2}_{e}C^{*}_{g}
-\Fu{0}_{\sigma,z}C_{e}|C_{g}|^2(1+|C_{e}|^2)
+\Fu{0}_{\sigma\dg,z}C^{2}_{g}C^{*}_{e}|C_{g}|^2\nl+\Fu{0}_{\sigma_{z},z}C_{g}
|C_{g}|^2(1+2|C_{e}|^2)-\Fu{0}_{I,z}C_{g}|C_{g}|^2.
\end{eqnarray}
\end{subequations}\end{widetext}
In this equation the noise function is given by
\begin{equation}
  z^{*}(t)=z^{*}_{\Lambda}(t)+\frac{\gamma \kappa}{4}e^{-\kappa t/2}\int_{0}^{t}
  e^{\kappa s/2}C_{g}(s)C^{*}_{e}(s)ds,
\end{equation}
where $z^{*}_{\Lambda}(t)$ is defined by the correlation
\begin{equation}
\tilde{ { E}}[{z}_{\Lambda}(t)z^{*}_{\Lambda}(s)]=\frac{\gamma
\kappa}{4}e^{-\kappa|t-s|/2}.
\end{equation}
This is generated by having $z^{*}_{\Lambda}(t)$ obey the following stochastic
differential equation,
\begin{equation} \label{zTLA}
 d_{t}z_{\Lambda}(t)=-\frac{\kappa}{2}z_{\Lambda}(t)+\frac{\kappa}{2}\rt{\gamma}\zeta(t),
\end{equation} with $z^{*}_{\Lambda}(0)$ being a Gaussian random variable (GRV) satisfying
\begin{equation}
{ E}[z_{\Lambda}(0)z^{*}_{\Lambda}(0)]=\frac{\kappa\gamma}{4}
 \end{equation}
Here $\zeta(t)$ is standard complex white noise \cite{Gar83} and
satisfies ${ E}[\zeta(t)\zeta^{*}(s)]=\delta(t-s)$.

\subsubsection{$0^{\rm th}$ Order Approximation}

For the simple memory function, $J=1$, which means $\F{0}_{z}(t)=\F{0}^{(j)}_{z}(t)$. The
$0^{\rm th}$ order approximation occurs when we assume the form for $\F{0}_{z}(t)$ in
\erf{pertbat0}. From \erf{simplememfunc2} this implies
\begin{equation}
  \F{0}_{z}(t)=\frac{\gamma}{2}(1-e^{-\kappa t/2})\hat{\sigma},
\end{equation} thus
\begin{subequations}
\begin{eqnarray}
\Fu{0}_{\sigma,z}(t)&=&\frac{\gamma}{2}(1-e^{-\kappa t/2}),\\
 \Fu{0}_{\sigma\dg,z}(t)&=&\Fu{0}_{\sigma_{z},z}(t)=\Fu{0}_{I,z}(t)=0.\hspace{0.8cm}
\end{eqnarray}
\end{subequations}

\subsubsection{$1^{\rm st}$ Order Approximation}

The $1^{\rm st}$ first order approximation occurs when we assume a form for
$\F{1}^{(j,k)}_{z}(t)$, by \erfs{pertbat1}{simplememfunc2} this means
\begin{equation}
  \F{1}_{z}(t)=\frac{\gamma}{2}
  (1-e^{-\kappa t/2})[\hat{\sigma},\F{0}_{z}(t)],
\end{equation} thus
\begin{subequations}
\begin{eqnarray}
\Fu{1}_{\sigma,z}(t)&=&{\gamma}
(1-e^{-\kappa t/2})\Fu{0}_{\sigma_{z},z}(t),\\
 \Fu{1}_{\sigma_{z},z}(t)&=&-\frac{\gamma}{2}
 (1-e^{-\kappa t/2})\Fu{0}_{\sigma\dg,z}(t),\\
\Fu{1}_{\sigma\dg,z}(t)&=&\Fu{1}_{I,z}(t)=0.
 \hspace{0.8cm}
\end{eqnarray}
\end{subequations}
The zero order functionals are found by applying the TLA operators to \erf{FjdotFinal},
giving
\begin{eqnarray}\label{FjdotFinalTLA}
d_t\F{0}_{z}(t)
 &=& \frac{\gamma \kappa}{4}\hat{\sigma}
 -\frac{\kappa}{2}\F{0}_{z}(t)
+z^{*}(t)[\hat{\sigma},\F{0}_{z}(t)]\nl-i
 [\frac{\Delta}{2}\hat{\sigma}_{z}+\frac{\chi}{2}
 \hat{\sigma}_{x},\F{0}_{z}(t)]-
 [\hat{\sigma}\dg\F{0}_{z}(t),\nl\F{0}_{z}(t)]
-\hat{\sigma}\dg\F{1}_{z}(t).
\end{eqnarray}
Using \erf{Fexpand} this gives the following four coupled
nonlinear equations \begin{widetext}
\begin{subequations}
\begin{eqnarray}
\label{FTLA1} d_t
\Fu{0}_{\sigma,z}(t)&=&\smallfrac{1}{4}\gamma\kappa-\frac{\kappa}{2}
\Fu{0}_{\sigma,z}(t)+{i}\Delta\Fu{0}_{\sigma,z}(t) -{i}\chi
\Fu{0}_{\sigma_{z},z}(t)+2z^{*}(t)\Fu{0}_{\sigma_{z},z}(t)
\nl+\Fu{0}^{2}_{\sigma,z}(t),
\\ d_{t}\Fu{0}_{\sigma\dg,z}(t)&=&-\frac{\kappa}{2}\Fu{0}_{\sigma\dg,z}(t)
+{i}\chi \Fu{0}_{\sigma_{z},z}(t)
-{i}\Delta\Fu{0}_{\sigma\dg,z}(t)+2\Fu{0}_{\sigma_{z},z}(t)[\Fu{0}_{I,z}(t)
-\Fu{0}_{\sigma_{z},z}(t)]
\nl-\Fu{0}_{\sigma\dg,z}(t)\Fu{0}_{\sigma,z}(t)-[\Fu{1}_{I,z}(t)
-\Fu{1}_{\sigma_{z},z}(t)],
\\d_{t}\Fu{0}_{\sigma_{z},z}(t)&=&-\frac{\kappa}{2}\Fu{0}_{\sigma_{z},z}(t)
+i\frac{\chi}{2}\Fu{0}_{\sigma\dg,z}(t)
-i\frac{\chi}{2}\Fu{0}_{\sigma,z}(t)-\Fu{0}_{\sigma,z}(t)
[\Fu{0}_{I,z}(t)-\Fu{0}_{\sigma_{z},z}(t)]\nl-z^{*}(t)
\Fu{0}_{\sigma\dg,z}(t)-\smallfrac{1}{2}\Fu{1}_{\sigma,z}(t),
\\\label{FTLA2}
d_{t}\Fu{0}_{I,z}(t)&=&-\frac{\kappa}{2}\Fu{0}_{I,z}(t)
-\smallfrac{1}{2}\Fu{1}_{\sigma,z}(t).
\end{eqnarray} \end{subequations}\end{widetext}
which can be solved in parallel with \erf{CoherentAmplitudes}.

\subsubsection{$2^{\rm nd}$ Order Approximation}

The $2^{\rm nd}$ order approximation occurs when we assume a form for
$\F{2}^{(j,k,l)}_{z}(t)$, by \erfs{pertbatn}{simplememfunc2} this means
\begin{equation}
  \F{2}_{z}(t)=\frac{\gamma}{2}(1-e^{-\kappa t/2})[\hat{\sigma},\F{1}_{z}(t)],
\end{equation}thus
\begin{subequations}\label{120}
\begin{eqnarray}
\Fu{2}_{\sigma,z}(t)&=&{\gamma}(1-e^{-\kappa t/2})\Fu{1}_{\sigma_{z},z}(t),\\
\Fu{2}_{\sigma_{z},z}(t)&=&-\frac{\gamma}{2}(1-e^{-\kappa t/2})\Fu{1}_{\sigma\dg,z}(t),\\
\Fu{2}_{\sigma\dg,z}(t)&=&\Fu{2}_{I,z}(t)=0.
 \hspace{0.8cm}
\end{eqnarray}
\end{subequations}
The zero order functionals are given by \erft{FTLA1}{FTLA2}, however we now need equations
for $\F{1}_{z}(t)$.
 The
fist order functionals are found applying TLA operators to \erf{1Fjdotfinal}. With a
memory function specified by \erf{simplememfunc2} we get
\begin{eqnarray}
 d_t
 \F{1}_{z}(t)&=&\frac{\gamma\kappa}{4}[\hat{\sigma},\F{0}_{z}(t)]
 -{\kappa}\F{1}_{z}(t)+z^{*}(t)\nl\times[\hat{\sigma},\F{1}_{z}(t)]
-i [\frac{\Delta}{2}\hat{\sigma}_{z}+\frac{\chi}{2}
 \hat{\sigma}_{x},\F{1}_{z}(t)]
 \nl-
[\hat{\sigma}\dg\F{1}_{z}(t),\F{0}_{z}(t)]-
[\hat{\sigma}\dg\nl\times\F{0}_{z}(t),\F{1}_{z}(t)]
 -\hat{\sigma}\dg\F{2}_{z}(t).
\end{eqnarray}
Using \erf{120} this turns into the four equations
\begin{widetext}\begin{subequations}
\begin{eqnarray}
 d_t \Fu{1}_{\sigma,z}(t)&=&\smallfrac{1}{2}\gamma\kappa
\Fu{0}_{\sigma_{z},z}(t)-{\kappa}\Fu{1}_{\sigma,z}(t)
+{i}\Delta\Fu{1}_{\sigma,z}(t) -{i}\chi
\Fu{1}_{\sigma_{z},z}(t)+2z^{*}(t)\Fu{1}_{\sigma_{z},z}(t)
\nl+2\Fu{0}_{\sigma,z}(t)\Fu{1}_{\sigma,z}(t),
\\
d_{t}\Fu{1}_{\sigma\dg,z}(t)&=&-{\kappa}\Fu{1}_{\sigma\dg,z}(t)
+{i}\chi \Fu{1}_{\sigma_{z},z}(t)
-{i}\Delta\Fu{1}_{\sigma\dg,z}(t)
+2\Fu{1}_{\sigma_{z},z}(t)[\Fu{0}_{I,z}(t)-\Fu{0}_{\sigma_{z},z}(t)]
\nl+2\Fu{0}_{\sigma_{z},z}(t)[\Fu{1}_{I,z}(t)-\Fu{1}_{\sigma_{z},z}(t)]
-[\Fu{1}_{\sigma\dg,z}(t)\Fu{0}_{\sigma,z}(t)+\Fu{0}_{\sigma\dg,z}(t)\Fu{1}_{\sigma,z}(t)]
\nl-\Fu{2}_{I,z}(t)+\Fu{2}_{\sigma_{z},z}(t),
\\
d_{t}\Fu{1}_{\sigma_{z},z}(t)&=&-\frac{\gamma\kappa}{4}\Fu{0}_{\sigma\dg,z}(t)-
{\kappa}\Fu{1}_{\sigma_{z},z}(t)+i\frac{\chi}{2}\Fu{1}_{\sigma\dg,z}(t)
-i\frac{\chi}{2}\Fu{1}_{\sigma,z}(t)
-\Fu{1}_{\sigma,z}(t)[\Fu{0}_{I,z}(t)\nl-\Fu{0}_{\sigma_{z},z}(t)]
-\Fu{0}_{\sigma,z}(t)[\Fu{1}_{I,z}(t)-\Fu{1}_{\sigma_{z},z}(t)]
-z^{*}(t)\Fu{1}_{\sigma\dg,z}(t)\nl-\smallfrac{1}{2}\Fu{2}_{\sigma,z}(t),\\
d_{t}\Fu{1}_{I,z}(t)&=&-{\kappa}\Fu{1}_{I,z}(t)-\smallfrac{1}{2}\Fu{2}_{\sigma,z}(t).
\end{eqnarray} \end{subequations}\end{widetext}

\begin{figure}
\includegraphics[width= .45\textwidth]{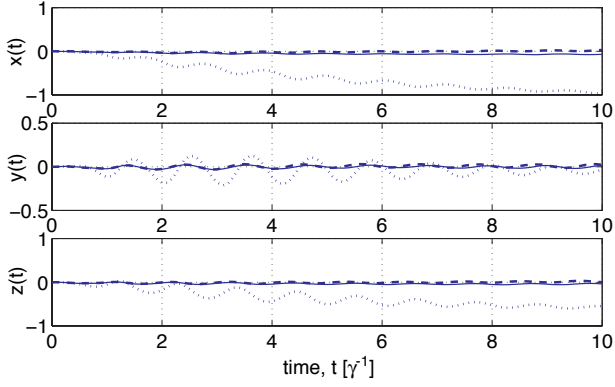}
\caption{\label{CoherentDiff} This figure depicts the difference
between the reduced state calculated form our perturbative
coherent non-Markovian SSE and the enlarged system method. The
dotted line corresponds to the $0^{\rm th}$ order perturbation,
the dashed is the $1^{\rm st}$ and the solid is the $2^{\rm nd}$.
Other details are as in Fig. \ref{Enlarged}.}
\end{figure}

To illustrate how accurate our perturbation method is, the
difference between the reduced state calculated via the enlarged
system method and the ensemble average from the coherent
non-Markovian SSE is plotted in Fig. \ref{CoherentDiff}. The
dotted line corresponds to the $0^{\rm th}$ order perturbation,
the dashed is the $1^{\rm st}$ and the solid is the $2^{\rm nd}$.
It is observed that the $1^{\rm st}$ and $2^{\rm nd}$ order
perturbation are a lot more accurate then the $0^{\rm th}$ order
perturbation. However, it can be seen that the $2^{\rm nd}$ order
perturbation is not necessarily more accurate than the $1^{\rm
st}$ order perturbation. This suggest that our perturbation method
is an asymptotic expansion rather than a convergent series.

\subsection{Quadrature Unravelling-TLA}

For the quadrature unravelling the actual non-Markovian SSE is
\begin{eqnarray}\label{QSSEOperatorFormTLA}
d_t\psizk{t}&=&\Big{[}
-i\frac{\Delta}{2}\hat{\sigma}_{z}-i\frac{\chi}{2}
 \hat{\sigma}_{x}- (\hat{\sigma}_{x}-\langle\hat{\sigma}_{x}\rangle_t)
~^{(0)}\hat{Q}_{z}(t)\nl +
\Big{\langle}(\hat{\sigma}_{x}-\langle\hat{\sigma}_{x}\rangle_t)
 ~^{(0)}\hat{Q}_{z}(t) \Big{\rangle}_t
\nl+z(t)(\hat{\sigma}-\langle\hat{\sigma}\rangle_t)\Big{
]}\psizk{t},
\end{eqnarray} and the noise function for the TLA is
\begin{equation}\label{QnoiseTLA}
z(t)=z_{\Lambda}(t)+\int_{0}^{t} \beta{(t-s)} \langle \hat{\sigma}_{x}\rangle_s  ds.
\end{equation}
Again the coherent case we can calculate the complex amplitude equation via applying the
state $\psizk{t}=C_{e}(t)\ket{e}+C_g(t)\ket{g}$ to \erf{QSSEOperatorFormTLA} and expanding
$\Q{0}_{z}(t)$ as
\begin{equation}\label{Qexpand}
  \Q{0}_{z}(t)=\sum_{m}\hat{m}\Qu{0}_{m,z}(t)
\end{equation} where $m=\{\sigma$, $\sigma\dg$, $\sigma_{z}$, $I\}$. This
results in a coupled set of differential equations for $C_{e}(t)$ and $C_g(t)$ that depend
on $\Qu{0}_{m,z}(t)$ and $z(t)$. In these equations the real-valued noise is given by
\begin{eqnarray}
  z(t)&=&z_{\Lambda}(t)+\frac{\gamma \kappa}{4}e^{-\kappa t/2}\int_{0}^{t}
  e^{\kappa s/2}[C_{g}(s)C^{*}_{e}(s)\nl+C^{*}_{g}(s)C_{e}(s)]ds,
\end{eqnarray}
where $z_{\Lambda}(t)$ is found by
\begin{equation}
\tilde{ {E}}[{z}_{\Lambda}(t)z_{\Lambda}(s)]=\frac{\gamma
\kappa}{4}e^{-\kappa|t-s|/2}.
\end{equation}
This is generated by
\begin{equation} \label{zTLAReal}
 d_{t}z_{\Lambda}(t)=-\frac{\kappa}{2}z_{\Lambda}(t)+\frac{\kappa}{2}\rt{\gamma}\xi(t)
 \end{equation}
with $z_{\Lambda}(0)$ being a GRV satisfying ${
E}[z_{\Lambda}(0)z^{*}_{\Lambda}(0)]={\kappa\gamma}/{4}$. Here
$\xi(t)$ is standard white noise and satisfies
$E[\xi(t)\xi^{*}(s)]=\delta(t-s)$ \cite{Gar83}.

\subsubsection{$0^{\rm th}$ Order Approximation}

The situation is greatly simplified with the memory function in \erf{simplememfunc}, as
$\beta(t,s)=\beta^{(j,cos)}(t,s)=\beta^{(j,cos)}(t,s)$, which in turn implies
$\Q{0}_{z}(t)= \Q{0}^{(j,\cos)}_{z}(t)=\Q{0}^{(j,\sin)}_{z}(t)$.

The $0^{\rm th}$ order approximation is to set
\begin{equation}
  \Q{0}_{z}(t)=\frac{\gamma}{2}(1-e^{-\kappa
  t/2})\hat{\sigma},
\end{equation} thus
\begin{subequations}
\begin{eqnarray}
\Qu{0}_{\sigma,z}(t)&=&\frac{\gamma}{2}(1-e^{-\kappa t/2}),\\
\Qu{0}_{\sigma\dg,z}(t)&=&\Qu{0}_{\sigma_{z},z}(t)=\Qu{0}_{I,z}(t)=0.
\end{eqnarray}
\end{subequations}

\subsubsection{$1^{\rm th}$ Order Approximation}

The first order approximation is to set
\begin{equation}
  ~\Q{1}_{z}(t)=\frac{\gamma}{2}(1-e^{-\kappa t/2})[\hat{\sigma},\Q{0}_{z}(t)]
\end{equation}thus
\begin{subequations}
\begin{eqnarray}
\Qu{1}_{\sigma,z}(t)&=&{\gamma}(1-e^{-\kappa t/2})\Qu{0}_{\sigma_{z},z}(t),\\
 \Qu{1}_{\sigma_{z},z}(t)&=&-\frac{\gamma}{2}(1-e^{-\kappa t/2})\Qu{0}_{\sigma\dg,z}(t),\\
\Qu{1}_{\sigma\dg,z}(t)&=&\Qu{1}_{I,z}(t)=0.
 \hspace{0.8cm}
\end{eqnarray}
\end{subequations}
 The $0^{\rm th}$ order functionals are found by applying TLA operators to
\erf{QjdotFinal}. With the simple memory function this gives
\begin{eqnarray}\label{QjdotFinalTLA}
  d_t \Q{0}_{z}(t)&=&\frac{\gamma\kappa}{4}\hat{\sigma}
   -\frac{\kappa}{2}\Q{0}_{z}(t)+z(t)[\hat{\sigma},\Q{0}_{z}(t)]
   \nl-i[\frac{\Delta}{2}\hat{\sigma}_{z}+\frac{\chi}{2}
 \hat{\sigma}_{x},\Q{0}_{z}(t)]\nl
 -[\hat{\sigma}_{x}\Q{0}_{z}(t),\Q{0}_{z}(t)]\nl -\hat{\sigma}_{x}\Q{1}_{z}(t) .
\end{eqnarray}
Using \erf{Qexpand} this gives,\begin{widetext}
\begin{subequations}
\begin{eqnarray}
\label{QTLA1}  d_t
\Qu{0}_{\sigma,z}(t)&=&\smallfrac{1}{4}\gamma\kappa-\frac{\kappa}{2}
\Qu{0}_{\sigma,z}(t)+{i}\Delta\Qu{0}_{\sigma,z}(t) -{i}\chi
\Qu{0}_{\sigma_{z},z}(t)+2z(t)\Qu{0}_{\sigma_{z},z}(t)
+\Qu{0}^{2}_{\sigma,z}(t) \nl-2\Qu{0}_{\sigma_{z},z}(t)
[\Qu{0}_{I,z}(t)+\Qu{0}_{\sigma_{z},z}(t)]-
\Qu{0}_{\sigma\dg,z}(t)\Qu{0}_{\sigma,z}(t)\nl
-[\Qu{1}_{I,z}(t)+\Qu{1}_{\sigma_{z},z}(t)] ,
\\
d_{t}\Qu{0}_{\sigma\dg,z}(t)&=&-\frac{\kappa}{2}\Qu{0}_{\sigma\dg,z}(t)
+{i}\chi \Qu{0}_{\sigma_{z},z}(t)
-{i}\Delta\Qu{0}_{\sigma\dg,z}(t)+2\Qu{0}_{\sigma_{z},z}(t)
[\Qu{0}_{I,z}(t)-\Qu{0}_{\sigma_{z},z}(t)]
\nl-\Qu{0}_{\sigma\dg,z}(t)\Qu{0}_{\sigma,z}(t)+\Qu{0}^{2}_{\sigma\dg,z}(t)
-\Qu{1}_{I,z}(t)+\Qu{1}_{\sigma_{z},z}(t),\\
d_{t}\Qu{0}_{\sigma_{z},z}(t)&=&-\frac{\kappa}{2}\Qu{0}_{\sigma_{z},z}(t)
+i\frac{\chi}{2}\Qu{0}_{\sigma\dg,z}(t)
-i\frac{\chi}{2}\Qu{0}_{\sigma,z}(t)-\Qu{0}_{\sigma,z}(t)
[\Qu{0}_{I,z}(t)-\Qu{0}_{\sigma_{z},z}(t)]\nl+\Qu{0}_{\sigma\dg,z}(t)
[\Qu{0}_{I,z}(t)+\Qu{0}_{\sigma_{z},z}(t)]-z(t)\Qu{0}_{\sigma\dg,z}(t)\nl-\smallfrac{1}{2}
[\Qu{1}_{\sigma,z}(t)-\Qu{1}_{\sigma\dg,z}(t)],\\
\label{QTLA2}
d_{t}\Qu{0}_{I,z}(t)&=&-\frac{\kappa}{2}\Qu{0}_{I,z}(t)-\smallfrac{1}{2}
[\Qu{1}_{\sigma,z}(t)+\Qu{1}_{\sigma\dg,z}(t)].
\end{eqnarray} \end{subequations}\end{widetext}
which can be solved in parallel with $C_{e}(t)$ and $C_g(t)$.

\begin{figure}
\includegraphics[width= .45\textwidth]{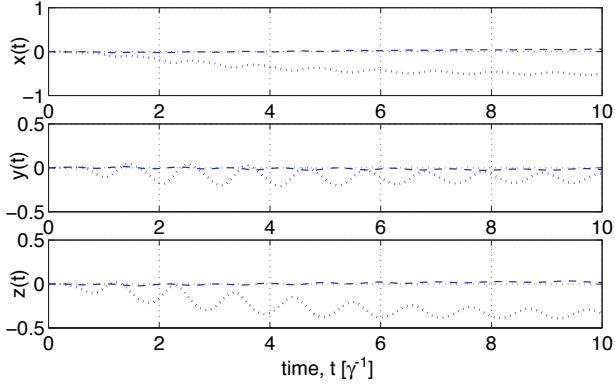}
\caption{\label{QuadratureDiff} This figure depicts the difference
between the reduced state calculated form our perturbative
quadrature non-Markovian SSE and the enlarged system method. The
dotted line corresponds to the $0^{\rm th}$  and the dashed is the
$1^{\rm st}$ order perturbation. Other details are as in Fig.
\ref{Enlarged}.}
\end{figure}

To illustrate how accurate our perturbation method is for the
quadrature unravelling. Fig. \ref{QuadratureDiff} shows the
difference between the reduced state calculated via the enlarged
system method and the ensemble average from the quadrature
non-Markovian SSEs for the $0^{\rm th}$ (dotted) and $1^{\rm st}$
(dashed) order perturbation. As in the coherent case we find the
$1^{\rm st}$ order perturbation is more accurate then the $0^{\rm
th}$.

\section{Post-Markovian perturbation} \label{YDGS}

In this section we extend the YDGS post-Markovian perturbation
\cite{YuDioGisStr99} to include the quadrature unraveling and
compare the post-Markovian method with our perturbation method.

The basis idea behind their perturbation method is to expand the
operators $ \f{0}_{z}(t,s)$ in powers of $(t-s)$ around the point
$t=s$ (this is why it is called the post Markovian perturbation).
That is
\begin{eqnarray}\label{fDpert}
  \f{0}_{z}(t,s)&=&\f{0}_{z}(s,s)+[\partial_{t}\f{0}_{z}(t,s)|_{t=s}](t-s)\nl
  +\smallfrac{1}{2}[\partial^{2}_{t}\f{0}_{z}(t,s)|_{t=s}](t-s)^{2}+...,
\end{eqnarray} where $\f{0}_{z}(s,s)=\hat{L}$.
To find the first order term we simply evaluate \erf{ComCoh} at $t=s$
\begin{eqnarray}\label{Dpert2}
  \partial_{t}\f{0}_{z}(t,s)|_{t=s}&=&-\frac{i}{\hbar}[\hat{H}_{\rm int}(s),\hat{L}]
  -[\hat{L}\dg\F{0}_{z}(s),\hat{L}]\nl-\hat{L}\dg[\hat{L},\F{0}_{z}(s)].
\end{eqnarray}
 Thus the functional $\F{0}_{z}(t)$ for this perturbation is given by
\begin{eqnarray}\label{Fdpert}
  \F{0}_{z}(t)
  &=&g_{0}(t)\hat{L}-g_{1}(t)\frac{i}{\hbar}[\hat{H}_{\rm
  int}(t),\hat{L}]\nl
  -\int_{0}^{t}\alpha(t-s)(t-s)[\hat{L}\dg\F{0}_{z}(s),\hat{L}]ds\nl-
   \int_{0}^{t}\alpha(t-s)(t-s)\hat{L}\dg[\hat{L},\F{0}_{z}(s)]ds,\hspace{1cm}
\end{eqnarray}where
\begin{eqnarray}\label{gs}
g_{0}(t)&=&\int_{0}^{t}\alpha(t-s)ds,\\
g_{1}(t)&=&\int_{0}^{t}\alpha(t-s)(t-s)ds.
\end{eqnarray} This equation can not be solved without the initial condition
$d_{t}\F{0}_{z}(0)$. However if we make the approximate $\F{0}_{z}(s)=
\int_{0}^{s}\alpha(s-u)\hat{L}du$, \erf{Fdpert} becomes
\begin{equation}\label{Fdpert2}
  \F{0}_{z}(t)=g_{0}(t)\hat{L}-g_{1}(t)\frac{i}{\hbar}[\hat{H}_{\rm
  int}(t),\hat{L}]-g_{2}(t)[\hat{L}\dg\hat{L},\hat{L}],
\end{equation} where
\begin{equation}
g_{2}(t)=\int_{0}^{t}\int_{0}^{s}\alpha(t-s)\alpha(s-u)(t-s)duds,
\end{equation} which can be solved.
The same could be done for the second order terms, but as well as making an approximation
for $\F{0}_{z}(s)$ we would need to approximate $d_{s}\F{0}_{z}(s)$. For the purpose of
this paper we will only go to first order.

To extend the idea to the quadrature case we Taylor expand the operator $\q{0}_{z}(t,s)$
in powers of $(t-s)$ around the point $t=s$. To find the first order term we simply
evaluate \erf{ComQuad} at $t=s$. With the approximation $\Q{0}_{z}(s)=
\int_{0}^{s}\beta(s-u)\hat{L}du$ we get
\begin{equation}\label{Qdpert}
  \Q{0}_{z}(t)=h_{0}(t)\hat{L}-h_{1}(t)\frac{i}{\hbar}[\hat{H}_{\rm
  int}(t),\hat{L}]-h_{2}(t)[\hat{L}_{x}\hat{L},\hat{L}].
\end{equation}
where
\begin{eqnarray}\label{hs}
h_{0}(t)&=&\int_{0}^{t}\beta(t-s)ds,\\
h_{1}(t)&=&\int_{0}^{t}\beta(t-s)(t-s)ds,\\
h_{2}(t)&=&\int_{0}^{t}\int_{0}^{s}\beta(t-s)\beta(s-u)(t-s)duds.
\end{eqnarray}

\begin{figure}
\includegraphics[width= .45\textwidth]{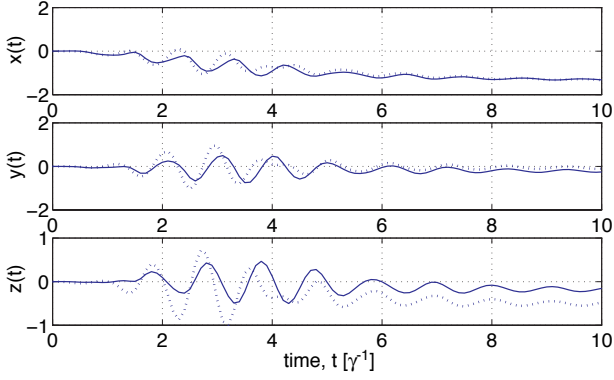}
\caption{\label{DpertDiff} This figure shows the difference between
 the reduced state calculated from YDGS
post-Markovian non-Markovian SSE method and the enlarged system
method, for both the coherent (dotted line) and quadrature (solid
line) unraveling. Other details are
 as in Fig. \ref{Enlarged}.}
\end{figure}

For the simple TLA system it is easy to generate these approximate expressions for
$\F{0}_{z}(t)$ and $\Q{0}_{z}(t)$ for all time, hence we can obtain solution to the
non-Markovian SSE. To compare YDGS post-Markovian non-Markovian SSE method with our
perturbation method, we again plot the difference between YDGS method (when 1000
trajectories where used) and the enlarged systems method. The results of this are shown in
Fig. \ref{DpertDiff}, where it is observed that YDGS first order perturbation has a
greater error than our perturbation method (Figs. \ref{CoherentDiff} and
\ref{QuadratureDiff}). This is perhaps not surprising, as the system we modelled has
$\kappa=1$, which implies it is very non-Markovian. Since one of the requirements of YDGS
perturbation method is for the environment to be close the Markovian regime one would
expect their method to fail in this regime.

In Ref. \cite{YuDioGisStr99} YDGS suggest an alternative
perturbation method. The functional operator $\F{0}_{z}(t)$, which
equals $\bar{O}_{z}(t)$ in their notation, is expanded by the
functional expansion \begin{widetext}\begin{eqnarray}
\bar{O}_{z}(t)&=&\bar{O}^{(0)}(t)+\int_{0}^{t}\bar{O}^{(1)}(t,v)z(v)dv
+\int_{0}^{t}\int_{0}^{t}\bar{O}^{(2)}(t,v_{1},v_{2})z(v_{1})z(v_{2})dv_{1}dv_{2}+...\nl+
\int_{0}^{t}...\int_{0}^{t}\bar{O}^{(n)}(t,v_{1},...,v_{n})z(v_{1})...z(v_{n})dv_{1}...dv_{n}+...,
\end{eqnarray}\end{widetext}
It can be shown that one can establish a set of coupled
differential equations for these operators provided $\alpha(t-s)$
is given by \erf{memoryFunctForm}. To truncated this perturbation
at $\bar{O}^{(n)}$ one has to assume a value $\bar{O}^{(n+1)}$. It
turns out that for all operators $\bar{O}^{(n)}$ other then $
\bar{O}^{(0)} $ the only reason the operators change from their
initial value $0$ at $t=0$ is if the assumed $\bar{O}^{(n+1)}$ is
nonzero. This suggest that this method is highly dependent on the
assumed value for $\bar{O}^{(n+1)}$.

\section{Conclusions} \label{Conclude}

 In this paper we presented a perturbation method for solving
the coherent and quadrature non-Markovian SSEs.  This perturbation
method is easily extended to any order and is not limited to the
post Markovian regime. However, the environment is restricted such
that it has a correlation function satisfying
\erf{memoryFunctForm}. As shown in Ref. \cite{SteIma96} most
non-Markovian environments can be simulated via this correlation
function with a relative small $J$. This suggest that this
perturbation method might be useful for simulating non-Markovian
evolution for $\rho_{\rm red}(t)$.

One appealing feature of this method is that it provides a
perturbative solution for $\rho_{\rm red}(t)$ which is positive by
definition. However there is another method, namely Imamo$\bar{\rm
g}$lu's enlarged system method \cite{Ima94,SteIma96}, which
provides a better solution for $\rho_{\rm red}(t)$. Imamo$\bar{\rm
g}$lu's enlarged system method requires fewer coupled differential
equations to solve and the only approximation comes in by a
truncation of the Hilbert space of the fictitious modes. As one
increases the basis size for these modes this method will converge
to the correct solution. By contrast, convergence has not been
shown for our method.

This does not mean that our method is useless, as the primary
interest in our method is not to simulate $\rho_{\rm red}(t)$, but
to simulate the non-Markovian SSEs. This is interesting as a
continuous in time interpretation of non-Markovian SSEs is not
clear. In Ref. \cite{GamWis02} we showed that these non-Markovian
SSE under standard quantum measurement theory do not have a
continuous measurement interpretation. However Loubenets in Ref.
\cite{Lou01,BarLou02} claimed that she has developed a new
framework for continuous quantum measurements in which
non-Markovian SSEs represent the evolution of a system state which
is continuously monitored.

Future work on this topic is to look into this question. Another
question that needs answering is whether it is possible to derive
non-Markovian SSE based on a discrete basis such as photon number.
We believe this question and the previous question will be
related. Finally, there is the possible application of our method
to strongly non-Markovian systems such as an atom laser
\cite{HopMoyColSav00} or photon emission in a photonic bad-gap
material \cite{Joh84,BayLamMol97}.

\appendix
\section{Derivation of $^{(0)}\hat{f}_{z}(t,t)=\hat{L}$}\label{AA}

To show that $^{(0)}\hat{f}_{z}(t,t)=\hat{L}$ we start by
 discretizing the functional derivative. We
divide the range $[0,t)$ into $N$ intervals of width $\Delta t$, so the change in
$\lpsizk{t}$ is
\begin{eqnarray}
  \delta \lpsizk{t}&=&\int_{0}^{t}\frac{\delta\lpsizk{t}}{\delta z^{*}(s)} \delta z^{*}(s)ds\nn\\&=&\sum_{i=0}^{N-1}\Delta t
  \Big{[} \frac{\partial\lpsizk{t_{N}}}{\partial z^{*}(t_{i})\Delta t} \Big{]}dz^{*}(t_{i}),
\end{eqnarray} thus
\begin{eqnarray}
  \frac{\delta}{\delta z^{*}(s)} \lpsizk{t}=\frac{\partial\lpsizk{t_{N}}}{\partial z^{*}(t_{i})\Delta t},
\end{eqnarray} if $s$ ($t_{i}$) is less than $t$ $(t_{N})$,
which is the only situation we are interested in, then taking the limit that $s\rightarrow
t$ ($t_{i}=t_{N-1}$) this becomes
\begin{equation} \label{step}
\lim_{s\rightarrow t}\frac{\delta\lpsizk{t}}{\delta z^{*}(s)}=\frac{\partial
   [\lpsizk{t_{N-1}}+\Delta t\partial_{t}\lpsizk{t_{N-1}}]}{\partial z^{*}(t_{N-1})\Delta t}.
\end{equation} Discretizing \erf{CLSSEFunctionalForm} we get
\begin{eqnarray}
\partial_{t}\lpsizk{t_{N-1}}&=&\Big{[}
\frac{-i}{\hbar}\hat{H}_{\rm
int}(t_{N-1})+z^{*}(t_{N-1})\hat{L}\nl-\hat{L}\dg\sum_{j=0}^{N-2}
\alpha{(t_{N-1}-t_{j})}\frac{\partial}{\partial
z^{*}(t_{j})}\Big{]}\nl\times\lpsizk{t_{N-1}}.
\end{eqnarray}
Substituting this into \erf{step} and using the fact that the
state at time $t_{N-1}$ only depends on the noise at time less
then $t_{N-1}$, we get the limit
\begin{equation}
 \frac{\delta\lpsizk{t}}{\delta z^{*}(t)}\rightarrow\hat{L}\lpsizk{t}.
\end{equation}
Thus by \erf{AnsatzO} $ \f{0}_{z}(t,t)=\hat{L}$.

\end{document}